\documentclass{article}

\usepackage{algorithm}
\usepackage[margin=30mm,includehead,includefoot]{geometry}
\usepackage[noend]{algpseudocode}
\usepackage{amsmath,amssymb,amsthm,amscd,color,comment}
\usepackage{autobreak}
\usepackage{color}
\usepackage{xcolor}
\usepackage{url}
\usepackage{blkarray}
\usepackage{jheppub}
\usepackage{empheq}
\usepackage{graphicx
}
\usepackage{booktabs, multirow}
\usepackage{subcaption}
\usepackage{bm}
\usepackage{enumerate}
\usepackage{mathtools}

\usepackage{tikz}
\usetikzlibrary{patterns}
\usetikzlibrary{arrows,shapes,positioning}
\usetikzlibrary{trees}
\usetikzlibrary{matrix,arrows} 				
\usetikzlibrary{positioning}				
\usetikzlibrary{calc,through}				
\usetikzlibrary{decorations.pathreplacing}  
\usepackage{pgffor}							
\usetikzlibrary{decorations.pathmorphing}	
\usetikzlibrary{decorations.markings}

\tikzset{
	compobj/.style={draw=black},
	binary/.style={draw=black, double distance=0.03cm},
	potgrav/.style={draw=black, dashed, dash pattern=on 3pt off 1.5pt,  line width=1pt},
	phigrav/.style={draw=blue, dashed, dash pattern=on 3pt off 1.5pt,  line width=1pt},
	Agrav/.style={draw=red, dashed, dash pattern=on 3pt off 1.5pt,  line width=1pt},
	sigmagrav/.style={draw=green, dashed, dash pattern=on 3pt off 1.5pt,  line width=1pt},
	radgrav/.style={decorate, draw=black, segment length=5pt, decoration={snake,amplitude=2pt} },
}
\tikzset{fermionnoarrow/.style={draw=black},	
        source/.style={draw=black, postaction={decorate}},
	    rgrav/.style={decorate, decoration={snake}, draw},
        bgrav/.style={dashed,dash pattern=on 3pt off 1.5pt,draw=black, line width=1pt, postaction={decorate}},
        cross/.pic = {
    \draw[rotate = 45] (-#1,0) -- (#1,0);
    \draw[rotate = 45] (0,-#1) -- (0, #1);
    }
     }

\newcommand{\dd}{{\rm d}}
\newcommand{\DD}{{\rm D}}
\newcommand{\ii}{{\rm i}}

\definecolor{green1}{HTML}{3D792A}
\definecolor{cyan1}{HTML}{37cdaa}
\definecolor{blue1}{HTML}{5d7ac4}
\definecolor{red1}{HTML}{d0482a}
\definecolor{purple1}{HTML}{845ea8}
\definecolor{orange1}{HTML}{e07229}

\usepackage{breqn}

\title{

Radiating Love: adiabatic tidal fluxes and modes up to next-to-next-to-leading post-Newtonian order
}

\author[a,b]{Manoj K. Mandal,}
\author[a,b]{Pierpaolo Mastrolia,}
\author[c,d]{Raj Patil,}
\author[c]{Jan Steinhoff}

\newcommand{\unipd}{Dipartimento di Fisica e Astronomia, Universit\`a degli Studi di Padova,
Via Marzolo 8, I-35131 Padova, Italy.}

\newcommand{\pdinfn}{INFN, Sezione di Padova,
Via Marzolo 8, I-35131 Padova, Italy.}

\affiliation[a]{\unipd}
\affiliation[b]{\pdinfn}
\affiliation[c]{Max Planck Institute for Gravitational Physics (Albert Einstein Institute), Am M{\"u}hlenberg 1, Potsdam 14476, Germany}
\affiliation[d]{Institut f{\"u}r Physik und IRIS Adlershof, Humboldt-Universit {\"a}t zu Berlin, Zum Großen Windkanal 2, D-12489 Berlin, Germany}

\emailAdd{manojkumar.mandal@pd.infn.it}
\emailAdd{pierpaolo.mastrolia@unipd.it}
\emailAdd{raj.patil@aei.mpg.de}
\emailAdd{jan.steinhoff@aei.mpg.de}

\abstract{
We present the analytical evaluation of the gravitational energy and angular momentum flux with tidal effects for inspiraling compact binaries, at the next-to-next-to-leading post-Newtonian (2PN) order, 
within the diagrammatic Effective Field Theory approach.
We first compute the stress-energy tensor for a binary system, which requires the evaluation of two-point Feynman integrals, up to two loops. Then we extract the multipole moments of the system, which we present for generic orbits in center-of-mass coordinates, and which are needed to evaluate the total gravitational energy and the angular momentum flux for generic orbits. 
Finally, we provide the expressions for gauge invariant quantities for circular orbits, such as the fluxes, mode amplitudes, and phase of the emitted gravitational wave. 
Our results are useful for updating previous theoretical studies, as well as related phenomenological analyses and waveform models.
}

\def\NNLO{N$^2$LO }
\def\NNNLO{N$^3$LO }

\allowdisplaybreaks

\begin{document}
\addtocontents{toc}{\protect\setcounter{tocdepth}{2}}

\begin{flushright}
\begingroup\footnotesize\ttfamily
 HU-EP-24/39-RTG
\endgroup
\end{flushright}

\maketitle

\section{Introduction}

The recent observations by the LIGO-Virgo-KAGRA collaboration \cite{LIGOScientific:2021djp} marked the dawn of the exploration in astronomy and cosmology through gravitational waves (GWs). The worldwide network of ground-based \cite{LIGOScientific:2014pky,VIRGO:2014yos,KAGRA:2020agh,Saleem:2021iwi,LIGOScientific:2016wof,Punturo:2010zza,Abac:2025saz} and forthcoming space-based GW detectors \cite{LISA:2017pwj} continues to grow, providing access to an ever-wider frequency band with increasing sensitivity. 
One of the most important GW sources we can detect are Neutron Star (NS) binaries  \cite{LIGOScientific:2017vwq,LIGOScientific:2017ync,LIGOScientific:2020aai}, which provide valuable insights into the physics of the dense nuclear matter in these stars. In such a binary system, the tidal interaction with a companion induces a quadrupole moment in the NS \cite{Flanagan:2007ix}. The imprint of these tidal interactions was observed in the GW signal GW170817 \cite{LIGOScientific:2017vwq}, leading to important constraints on the neutron star equation of state (EOS) \cite{LIGOScientific:2018hze,LIGOScientific:2018cki,Chatziioannou:2020pqz,Pradhan:2022rxs}.

In this article, we use Effective Field Theory (EFT) techniques~\cite{Goldberger:2004jt} to analyze the binary's inspiral, i.e., when the binary components are moving at nonrelativistic velocities and the orbital separation is large.
In this regime, we can use a perturbative approach involving a series expansion in powers of $v/c$, where $v$ is the orbital velocity of the binary and $c$ is the speed of light. 
The virial theorem requires that the kinetic energy be on average $-1/2$ times the potential energy of a bound state system.
Therefore, we can perform a \emph{post-Newtonian} (PN) analysis, which involves an expansion in two perturbative parameters: $v/c$ and $G_N$, where $G_N$ is Newton's constant. 
Terms of order $(v/c)^{n}$ are said to be of $(n/2)$PN order.
The PN analysis of binary dynamics can be divided into two sectors, namely the conservative sector, where the emitted radiation is neglected and the orbital separation does not decrease, and the radiative sector, where the emitted radiation carries away energy and momentum. 
At higher PN orders, these sectors may mix due to tail effects resulting from radiation scattered back by the orbital background curvature and eventually interacting again with the orbital dynamics (see, e.g., Ref.~\cite{Blanchet:2013haa}).
Using the EFT approach, we can determine any observable quantity at any given PN order. By using modern diagrammatic EFT methods, first proposed in Ref.~\cite{Goldberger:2004jt}, and modern integration methods \cite{Kol:2013ega,Foffa:2016rgu}, the problem is transformed into the determination of scattering amplitudes.
These amplitudes can be obtained systematically by calculating the corresponding Feynman diagrams. See, e.g., Refs.~\cite{Porto:2016pyg,Levi:2018nxp,Goldberger:2022ebt} for reviews. For similar computations using traditional approaches, see Ref.~\cite{Blanchet:2013haa}.

The detection and interpretation of gravitational waves depends critically on accurate waveform models as input to the data analysis. 
Two key components of these models are the Hamiltonian, which describes the conservative dynamics of the binary source, and the fluxes, which describe its radiative dynamics.
The state-of-the-art for the conservative Hamiltonian of the point-particle sector is at the 4PN order, computed using classical GR methods 
\cite{Damour:2014jta,Jaranowski:2015lha,Bernard:2015njp,Bernard:2016wrg,Damour:2016abl}, and using the EFT diagrammatic approach~\cite{Foffa:2012rn,Foffa:2019rdf,Foffa:2016rgu,Foffa:2019yfl,Blumlein:2020pog}. Preliminary results are also available at higher orders \cite{Foffa:2019hrb,Blumlein:2021txe,Porto:2024cwd,Blumlein:2021txj,Blumlein:2020znm,Bini:2019nra}.
For the radiative sector, the state-of-the-art for the phase is at 4.5PN order~\cite{Blanchet:2023bwj}, 
and for the flux and the quadrupole moment is at 4PN order Ref.~\cite{Blanchet:2023sbv}, using classical GR methods.
EFT techniques have also been used in the context of radiative dynamics to derive the energy flux at 3PN order~\cite{Amalberti:2024jaa}.

In this article we are interested in binary sources with tidally deformed compact objects. 
For this case, the state-of-the-art Hamiltonian up to \NNNLO (3PN) was recently derived using the EFT diagrammatic approach in Ref.~\cite{Mandal:2023hqa}, where a renormalization of the post-adiabatic Love number was observed (see also the related renormalization of the mass-quadrupole moment in Ref.~\cite{Goldberger:2009qd}).
For the radiative dynamics, \NNLO (2PN) corrections were first considered in Ref.~\cite{Henry:2020ski,Henry:2020skiErr}, using classical GR methods.\footnote{ 
Notably, the original results presented in the current article, anticipated to the authors of~\cite{Henry:2020ski}, 
helped them to identify a mistake and correct their calculations, 
as documented in the recent update in Ref.~\cite{Henry:2020skiErr}.}

In this work, we compute the stress-energy tensor of a binary system with tidally deformed compact objects, up to 2PN order, using the EFT diagrammatic approach and the multi-loop Feynman calculus. 
\textit{We present
the instantaneous energy flux, and the instantaneous angular momentum flux for generic orbits, and for the specialized case of circular orbits we present the total energy flux, total angular momentum flux, the phase, and the dominant quadrupolar mode amplitudes $(2,\pm 2)$  of the emitted gravitational wave, up to \NNLO (2PN)}. 
For self-consistency, we check the conservation of all components of the stress-energy tensor up to 2PN order.

The paper is organized as follows.
In Section~\ref{sec_EFT_descrip}, we review
the description of tidally-interacting binaries in the EFT formalism, and the procedure to compute the conservative and dissipative effects.  
In Section~\ref{sec_com_algo}, we present the algorithm used to compute the multipole moments and the procedure of gauge-transforming the radiative sector.
Our main results, the multipole moments, the energy and angular momentum fluxes, the mode amplitude, and the phase of the emitted gravitational waves, is presented in Section~\ref{sec_results}. 
Finally, we present our conclusions and avenues for future work in Section~\ref{sec_Conclusion}. 
In the appendix \ref{app_non_tidal}, we present non-tidal Lagrangian and multipole moments, and in appendix \ref{app_tidal} we present the Lagrangian for the tidal sector.
This work is supplemented with four ancillary files: 
\texttt{Stress\_Energy\_Tensor.m}, containing the analytic expression of the full stress-energy tensor up to 2PN including tidal effects, \texttt{Multipole\_Moments.m}, containing the analytic expression of the all the multipole moments given in section \ref{sec_results}, \texttt{Fluxes.m} that contains the expression for energy and angular momentum flux, and \texttt{Lagrangian.m} that contains the analytic expression of the Lagrangian presented in appendix \ref{app_non_tidal} and \ref{app_tidal}.

{\it Notation --}~ We work in $d+1$ spacetime dimension. The mostly negative signature for the metric is employed. Bold-face characters are used for three-dimensional variables, and normal-face font for four-dimensional variables. The subscript ${(a)}$ labels the binary components on all the corresponding variables, like their position $\bm{x}_{(a)}$. An overdot denotes the time derivative, e.g., $\bm{v}_{(a)}=\dot{\bm{x}}_{(a)}$ is the velocity, $\bm{a}_{(a)}=\ddot{\bm{x}}_{(a)}$ the acceleration. The separation between two objects is denoted by $\bm{r}=\bm{x}_{(1)}-\bm{x}_{(2)}$, with absolute value $r=|\bm{r}|$ and the unit vector along the separation is $\bm{n}=\bm{r}/r$. The multi-index notation is given by $x^L = x^{i_1}x^{i_2}\cdots x^{i_l}$ and for multipole moments $\mathcal{I}^{L}=\mathcal{I}^{i_1 i_2 \cdots i_l}$ and $\mathcal{J}^{L}=\mathcal{J}^{i_1 i_2 \cdots i_l}$. \texttt{STF} refers to symmetric trace-free components of the free indices and $A^{[ab]}=1/2(A^{ab}-A^{ba})$ is the notation for anti-symmetrization.

\section{EFT for tidally deformed compact objects}
\label{sec_EFT_descrip}

In the context of two compact objects in a bound state, there are three primary length scales to consider: the Schwarzschild radius of each compact object ($R_s$); the radius of the orbit ($r$); and the wavelength of the gravitational waves emitted ($\lambda$). Assuming that the velocities of the compact objects are much smaller than the speed of light and that the compact objects are widely separated, the background spacetime can be approximated as Minkowski flat, described by the metric $g_{\mu\nu}=\eta_{\mu\nu}+h_{\mu\nu}$, where $h_{\mu\nu}$ represents the perturbations due to the gravitational interaction between the compact objects. This setup gives rise to a hierarchy of scales:
\begin{align}
\lambda\gg r\gg R_s\, .
\end{align}
As we are only interested in the long-distance physics at the scales of $\lambda$, we first decompose the graviton fields as $h_{\mu\nu} = H_{\mu\nu} + \bar{h}_{\mu\nu}$ \cite{Goldberger:2004jt}, where the short distance modes (potential gravitons) $H_{\mu\nu}$ scale as $(k_0,\textbf{k})\sim(v/r,1/r)$ and long-distance modes (radiation gravitons) $\bar{h}_{\mu\nu}$ scale as $(k_0,\textbf{k})\sim(v/r,v/r)$.

The dynamics of the gravitational field ($g_{\mu\nu}$) is given by the Einstein-Hilbert action along with a gauge fixing term,
\begin{align}\label{eq_action_EH}
S_{\text{EH}} = -\frac{c^4}{16 \pi G_N} \int d^4x \sqrt{g} ~R[g_{\mu\nu}] 
+\frac{c^4}{32 \pi G_N }\int d^4x \bar{\Gamma}_\mu\bar{\Gamma}_\nu \bar{g}^{\mu\nu} \, ,
\end{align}
where, $G_N$ is the Newton's constant,  $R$ is the Ricci scalar, and $g$ is the determinant of the $g_{\mu\nu}$.
Here we use a background field gauge for the potential modes so that the EFT obtained after integrating them is gauge invariant for the radiation modes. This gauge is defined by
\begin{align}
	\bar{\Gamma}_\mu = \bar{g}^{\alpha\beta} \bar{\nabla}_\alpha H_{\beta\mu} - \frac{1}{2} \bar{g}^{\alpha\beta} \bar{\nabla}_\mu H_{\alpha\beta}\, ,
\end{align}
where $\bar{\nabla}_\alpha$ is defined on the background metric $\bar{g}_{\mu\nu} = \eta_{\mu\nu} + \bar{h}_{\mu\nu}$, and $\bar{g}$ is the determinant of the $\bar{g}_{\mu\nu}$. For computational convenience we also write the potential gravitons as $H_{\mu\nu}=g_{\mu\nu}-\eta_{\mu\nu}$ where then $g_{\mu\nu}$ is written in terms of the Kaluza-Klein (KK) fields : a scalar $\bm{\phi}$, a three-dimensional vector $\bm{A}$ and a three-dimensional symmetric rank two tensor $\bm{\sigma}$~\cite{Kol:2007bc,Kol:2007rx}, given by
\begin{equation}
g_{\mu\nu} = 
\begin{pmatrix}
e^{2\bm{\phi}/c^2} \,\,\, & -e^{2\bm{\phi}/c^2} {\bm{A}_j}/{c^2}\\
-e^{2\bm{\phi}/c^2} {\bm{A}_i}/{c^2} \,\,\,\,\,\, & -e^{-2\bm{\phi}/((d-2)c^2)}\bm{\gamma}_{ij}+e^{2\bm{\phi}/c^2} {\bm{A}_i}{\bm{A}_j}/{c^4}  
\end{pmatrix}\,,
\quad {\rm with} \quad
\bm{\gamma}_{ij}=\bm{\delta}_{ij}+\bm{\sigma}_{ij}/c^2 \,.
\end{equation}

To describe adiabatically tidally deformed objects, in particular objects with quadrapolar deformation that is locked to the external tidal field, we use the action \cite{Hinderer:2007mb,Binnington:2009bb,Damour:2009vw},
\begin{align}\label{eq_action_pp}
S_{\text{pp}} = \sum_{a=1,2} \int d\tau \left( - m_{(a)}c z_{(a)} + \frac{z_{(a)}\lambda_{(a)}}{4}  E_{(a)\mu\nu}E_{(a)}^{\mu\nu} \right)\, ,
\end{align}
where, $u^\mu=\dd x^\mu/\dd\tau$ is the four-velocity, $E_{(a)\mu\nu}= - c^2 R_{\mu\alpha\nu\beta}u_{(a)}^\alpha u_{(a)}^\beta/z_{(a)}^2$ is the gravitoelectric field, which is the relativistic analogue of the Newtonian external tidal field, $z_{(a)}=\sqrt{u_{(a)}^2}$ is the redshift factor, and $\tau$ is the proper time, related to the coordinate time $t$ as $\dd \tau=c~\dd t$.

We can now obtain the effective action for the binary and radiation gravitons by integrating out the potential gravitational degrees of freedom as follows \cite{Goldberger:2004jt} 
\begin{align}
\label{eq_tot_eff_action}
\exp \left[{\ii \, \int \dd t ~( \mathcal{L}_{\text{eff}}} + \Gamma_{\text{eff}}[\bar{h}] )\right] = \int \DD \bm{\phi} \, \DD \bm{A}_i \, \DD \bm{\sigma}_{ij}\, \exp[\ii \, (S_{\text{EH}}+S_{\text{pp}})] \, ,
\end{align}
where the Einstein-Hilbert action $S_{\rm EH}$ is given by Eq.~\eqref{eq_action_EH} and the point-particle action $S_{\rm pp}$ is given by Eq.~\eqref{eq_action_pp}. Here, $\mathcal{L}_{\rm eff}$ is the effective Lagrangian that describes the conservative dynamics of the binary. This is further decomposed as 
\begin{equation}
 \mathcal{L}_{\rm eff} = \mathcal{K}_{\rm eff} - \mathcal{V}_{\rm eff}\, ,
\end{equation}
where, $\mathcal{K}_{\rm eff}$ is the kinetic term and $\mathcal{V}_{\rm eff}$ is the effective contribution due to gravitational interactions between the two objects. The dissipative dynamics of the binary is encoded in the effective one point function of the radiation gravitons. This is given by
\begin{equation}
\label{eq_eff_one_pt_function}
  \Gamma_{\text{eff}}[\bar{h}] = -\frac{1}{2}\int d^d\textbf{x} \mathcal{T}_{\text{eff}}^{\mu\nu} \bar{h}_{\mu\nu} ,
\end{equation}
where the effective stress-energy tensor $\mathcal{T}_{\text{eff}}^{\mu\nu}$ encodes the information of the multipole moments of the binary and thus the fluxes and gravitational waveform emitted by it.

The terms that are obtained after performing the explicit integral are collectively denoted by the potential $\mathcal{V}_{\rm eff}$ and the effective one point function $\Gamma_{\text{eff}}[\bar{h}]$. These terms are computed by summing over the connected Feynman diagrams without graviton loops, as shown below,
\begin{align}
\mathcal{V}_{\text{eff}} = \ii \,
\lim_{d\rightarrow 3} 
\int \frac{\dd^d \bm{p}}{(2\pi)^d}~
e^{\ii \, \bm{p}\cdot (\bm{x}_{(1)}-\bm{x}_{(2)})}
\, 		
\parbox{25mm}{
	\begin{tikzpicture}[line width=1 pt,node distance=0.4 cm and 0.4 cm]
	\coordinate[label=left: ] (v1);
	\coordinate[right = of v1] (v2);
	\coordinate[right = of v2] (v3);
	\coordinate[right = of v3] (v4);
	\coordinate[right = of v4, label=right: \tiny$(2)$] (v5);
	\coordinate[below = of v1] (v6);
	\coordinate[below = of v6, label=left: ] (v7);
	\coordinate[right = of v7] (v8);
	\coordinate[right = of v8] (v9);
	\coordinate[right = of v9] (v10);
	\coordinate[right = of v10, label=right: \tiny$(1)$] (v11);
	\fill[black!25!white] (v8) rectangle (v4);
	\draw[fermionnoarrow] (v1) -- (v5);
	\draw[fermionnoarrow] (v7) -- (v11);
	\end{tikzpicture}
}
\, ,
\label{eq:effective_lagrangian}
\end{align}
\begin{align}
\Gamma_{\text{eff}}[\bar{h}] = -\ii \,
\lim_{d\rightarrow 3} \frac{1}{2} \int d^3\textbf{x}
\int \frac{\dd^d \bm{p}}{(2\pi)^d}~
e^{\ii \, \bm{p}\cdot (\bm{x}_{(1)}-\bm{x}_{(2)})}
\, 		
\parbox{25mm}{
	\begin{tikzpicture}[line width=1 pt,node distance=0.4 cm and 0.4 cm]
	\coordinate[label=left: ] (v1);
	\coordinate[right = of v1] (v2);
	\coordinate[right = of v2] (v3);
	\coordinate[right = of v3] (v4);
	\coordinate[right = of v4, label=right: \tiny$(2)$] (v5);
	\coordinate[below = of v1] (v6);
	\coordinate[below = of v6, label=left: ] (v7);
	\coordinate[right = of v7] (v8);
	\coordinate[right = of v8] (v9);
	\coordinate[right = of v9] (v10);
	\coordinate[right = of v10, label=right: \tiny$(1)$] (v11);
	\fill[black!25!white] (v8) rectangle (v4);
	\draw[fermionnoarrow] (v1) -- (v5);
	\draw[fermionnoarrow] (v7) -- (v11);
    \draw[radgrav] (1.2,-0.4)--(1.8,-0.4);
	\end{tikzpicture}
}
\, ,
\label{eq:effective_lagrangian}
\end{align}
where $\bm{p}$ is the linear momentum transferred between the two bodies. For details on the computation of the effective potential $\mathcal{V}_{\rm eff}$, refer to our previous article \cite{Mandal:2022nty}, and references therein.
In this article, we aim to compute the effective one-point function $\Gamma_{\text{eff}}[\bar{h}]$ using the techniques of effective field theories and multi-loop Feynman calculus, and extract the multipole moments, energy and angular momentum fluxes and gravitational waveform.


\section{Computational Algorithm}
\label{sec_com_algo}

\begin{table}
	\begin{subtable}[H]{0.49\textwidth}
    \centering
		\begin{tabular}{|c|c|c|c|}
			\hline
			Order                & Diagrams            & Loops       & Diagrams  \\ \hline
0PN                  & 1                   & -      & 1   \\ \hline
0.5PN                  & 1                   & -      & 1   \\ \hline
\multirow{2}{*}{1PN} & \multirow{2}{*}{3}  & 0    & 2 \\ \cline{3-4} 
                     &                     & -  & 1\\ \hline
\multirow{2}{*}{1.5PN} & \multirow{2}{*}{5}  & 0    & 4 \\ \cline{3-4} 
                     &                     & -  & 1 \\ \hline
\multirow{3}{*}{2PN} & \multirow{3}{*}{18} & 1    & 6 \\ \cline{3-4} 
                     &                     & 0 & 11 \\ \cline{3-4} 
                     &                     & - & 1 \\ \hline
		\end{tabular}
		\caption{Point particle sector}
		\label{tbl_no_diag_non_spinning}
	\end{subtable}
	\begin{subtable}[H]{0.49\textwidth}
    \centering
		\begin{tabular}{|c|c|c|c|}
			\hline
			Order                & Diagrams            & Loops       & Diagrams  \\ \hline
0PN                  & 1                   & 0      & 1   \\ \hline
0.5PN                  & 1                   & 0      & 1   \\ \hline
\multirow{2}{*}{1PN} & \multirow{2}{*}{6}  & 1    & 3 \\ \cline{3-4} 
                     &                     & 0  & 3 \\ \hline
\multirow{2}{*}{1.5PN} & \multirow{2}{*}{9}  & 1    & 6 \\ \cline{3-4} 
                     &                     & 0  & 3 \\ \hline
\multirow{3}{*}{2PN} & \multirow{3}{*}{68} & 2    &  25 \\ \cline{3-4} 
                     &                     & 1 & 37 \\ \cline{3-4} 
                     &                     & 0  & 6 \\ \hline
		\end{tabular}
		\caption{Adiabatic sector}
		\label{tbl_no_diag_tidalEQ}
	\end{subtable}
   
	\label{tbl_no_diag}
	\caption{Number of Feynman diagrams contributing different sectors of stress-energy tensor. Here - refers to disconnected diagrams that contribute due to single object graviton emission.
 }
\end{table}

\subsection{Effective stress-energy tensor}

To obtain the stress-energy tensor from the diagrammatic approach as shown in equation \eqref{eq:effective_lagrangian}, we begin by generating all the relevant generic topologies for one-point graviton emission contributing at different orders of $G_N$. 
To compute the terms proportional to $G_N^l$ where $l = 1, 2, ..., n + 1$, and we consider all the topologies at $l-1$ loops contributing to the specific order $l$.
So, for the computation of the 2PN adiabatic tidal stress-energy tensor, we generate all the topologies until the order $G_N^2$ (2-loop)
using \texttt{QGRAF}~\cite{NOGUEIRA1993279}. 
There is 1 disconnected topology that contributes to radiation from a single worldline, 2 topology at tree-level ($G_N$), 8 topologies at one-loop ($G_N^2$), and 48 topologies at two-loop ($G_N^3$).
Then we dress these topologies with the KK fields $\bm{\phi}$, $\bm{A}$ and $\bm{\sigma}$ and Feynman rules derived from the PN expansion of action given in \eqref{eq_action_EH} and \eqref{eq_action_pp}, to obtain all the Feynman diagrams that contribute to the given order of $G_N$ and $v$ depending on the specific perturbation order. 
The number of diagrams that contribute at particular order in $1/c$ and of particular loop topology are given in table \ref{tbl_no_diag_non_spinning} and \ref{tbl_no_diag_tidalEQ}\footnote{While considering the tidal effects, we count only the representative Feynman diagrams, where the tides can contribute from any of the worldline graviton interaction vertex present in the diagram. 
Additionally, the diagrams, which can be obtained from the change in the label $1\leftrightarrow 2$, are not counted as separate diagrams.}. For computing these diagrams, we use an in-house code that uses tools from \texttt{EFTofPNG}~\cite{Levi:2017kzq} and \texttt{xTensor}~\cite{xAct}, for the tensor algebra manipulation. 
The general structure of the one point function in the momentum space is of the form
\begin{equation}
    \Gamma_{G}^{(l)} =N^{\mu_1,\mu_2,\cdots}_{C~~~\nu_1,\nu_2,\cdots}
    \big(x_{(a)},q,\bar{h},\cdots\big) \int_p e^{\ii p_\mu (x_{(1)}-x_{(2)})^\mu} N_{F~~~\mu_1,\mu_2,\cdots}^{\alpha_1,\alpha_2,\cdots} (p,q) \prod_{i=1}^l \int_{k_i}  
    \frac{N_{M~~~\alpha_1,\alpha_2,\cdots}^{\nu_1,\nu_2,\cdots} (k_i)}{\prod_{\sigma \in G} D_\sigma (p, k_i,q)}\, ,
\end{equation}
    where, 
\begin{enumerate}
    \item  $N_C$ is a tensor polynomial depending on the world-line coordinates ($x_{(a)}^\mu$), and the radiation graviton ($\bar{h}_{\mu\nu}$) and its  momentum ($q$),
    \item $p$ is the momentum transfer between the sources (Fourier momentum), 
    \item $N_F$ is the tensor polynomial built out of momenta $p$ and $q$,
    \item $k_i$ are the loop momentums, 
    \item $D_\sigma$ denotes the set of denominators corresponding to the internal lines of $G$,
    \item $N_M$ stands for a tensor polynomial built out of loop momenta $k_i$.
\end{enumerate}
Since the momentum of the potential graviton scales as 
$(k_0,\textbf{k})\sim(v/r,1/r)$ and radiation gravitons $\bar{h}_{\mu\nu}$ scales as $(q_0,\textbf{q})\sim(v/r,v/r)$, the ratio $\textbf{q}/\textbf{k}\approx 1/c$ and hence can be expanded in the PN approximation. Hence, we take the soft limit in the denominators that can be expanded as a series in $q$ as,
\begin{align}
    \frac{1}{(\textbf{k}+\textbf{q})^2}
    =\frac{1}{\textbf{k}^2} \Bigg(1-2 \frac{\textbf{k}^\mu \textbf{q}_\mu}{\textbf{k}^2}- \frac{\textbf{q}^2}{\textbf{k}^2} + \cdots\Bigg)\, ,
\end{align}
so that only $N_C$ depends on $q$.
Upon the series expansion, the momentum space Feynman diagrams are mapped onto two-point massless integrals \cite{Foffa:2016rgu}, as graphically shown in Fig.~\ref{fig_relation_bet_diags}. We use \texttt{LiteRED} \cite{Lee:2013mka}, for the integration-by-parts reduction, to linearly express these integrals in terms of a minimal set of Master Integrals (MI). In the considered case, we obtain 1 MI at one loop, and 2 MIs at two loops.
Once the exact expressions of the master integrals are substituted in, we evaluate a Fourier transform over the transferred momentum $q$, to finally obtain the effective one-point function $\Gamma_{\text{eff}}[\bar{h}]$. The details of the algorithm and the expressions for the master integrals and Fourier integrals can be found in Ref.~\cite{Mandal:2022nty}.
\begin{figure}[t]
	\centering
	\begin{tikzpicture}[line width=1 pt, scale=0.4]
	\begin{scope}[shift={(-7,0)}]
	\filldraw[color=gray!40, fill=gray!40, thick](0,0) rectangle (3,3);
	\draw (-1.5,0)--(4.5,0);
	\draw (-1.5,3)--(4.5,3);
    \draw[radgrav] (3,1.5)--(5,1.5);
	\node at (1.5,6.5) {Gravity};	
	\node at (1.5,5.3) {Diagrams};	
    \node at (-5,1.5) {~};	
	\end{scope}
	\begin{scope}[shift={(0,0)}]
	\node at (0,6) {$\longleftrightarrow$};
	\node at (0,1.5) {$\equiv$};
	\end{scope}
	\begin{scope}[shift={(5,0)}]
	\filldraw[color=gray!40, fill=gray!40, thick](0,1.5) circle (1.5);
	\draw (0,3)--(0,4);
	\draw (0,0)--(0,-1);	
	\node at (0,6.5) {Multi-loop};				
	\node at (0,5.3) {Diagrams};	
    \draw[black,fill=black] (1.5,1.5) circle (1ex);
	\node at (2.3,1.5) {$^{\mu\nu}$};	
    \node at (5,1.5) {$\otimes~\bar{h}_{\mu\nu}\Big|_{k\rightarrow 0}$};	
    \end{scope}		
	\end{tikzpicture}
	\caption{
	}
	\label{fig_relation_bet_diags}
\end{figure}

This algorithm leads us to the result of the 2PN effective one point function in the momentum space that depends on the binary variables $x_{(a)}$, its time derivatives, and $\lambda_{(a)}$, the radiation gravitons $\bar{h}$ and its momentum $q$. 
Here one can easily read off the stress-energy tensor of the binary in momentum space
\begin{align}
    \mathcal{T}_{\text{eff}}^{\mu\nu}(t,\textbf{q}) =\int d^3\textbf{x} ~e^{\ii \textbf{q}\cdot\textbf{x}}~\mathcal{T}_{\text{eff}}^{\mu\nu}(t,\textbf{x})= \sum^{\infty}_{n=0} \frac{\ii^n}{n!} 
    \bigg( \int d^3\textbf{x} ~\mathcal{T}_{\text{eff}}^{\mu\nu}(t,\textbf{x}) ~\textbf{x}^{i_1}\cdots\textbf{x}^{i_n}\bigg) \textbf{q}_{i_1}\cdots\textbf{q}_{i_n} \, ,
\end{align} 
From this, we can extract all the relevant integrals of the stress-energy that contribute to different multipole moments as can be seen from section \ref{sec_sub_SE_to_Multipoles}. 

\subsection{Conservation of stress-energy tensor}
Since we use the background field gauge for the potential gravitons, the resulting stress-energy tensor must satisfy the conservation condition $\partial_\mu \mathcal{T}^{\mu\nu}=0$. This property allows us to perform stringent self-consistency checks on the components of the stress-energy tensor, specifically, $\mathcal{T}^{00}$, $\mathcal{T}^{0i}$, and $\mathcal{T}^{ij}$ ensuring their validity up to the 2PN order.
To verify these components, we proceed step-by-step using different moment relations:
First, we use the $\mathcal{T}^{00}$ component up to 1PN order in the following moment equation:
\begin{align}
        \int d^3\textbf{x}~ \mathcal{T}^{ij} &=\frac12 \frac{1}{c^2} \frac{d^2}{dt^2} \int d^3\textbf{x}~ \mathcal{T}^{00} x^ix^j \, .
\end{align}
This serves as a consistent check for the spatial component $\mathcal{T}^{ij}$ up to 2PN order. Next, we use the following moment equation:
\begin{align}
    \int d^3 \textbf{x}~ \mathcal{T}^{0 i} &= \frac{1}{c} \frac{d}{dt} \int d^3\textbf{x}~ \mathcal{T}^{00} x^i \, ,
    \label{eq:stress-energytencsor_chk-2}
\end{align}
where we use $\mathcal{T}^{00}$ component up to 1.5PN order, which allows us to verify the consistency of $\mathcal{T}^{0 i}$ up to 2PN order. 
Finally, to ensure the accuracy of the $\mathcal{T}^{00}$ component up to the 2PN order, we need the $\mathcal{T}^{0 i}$ component up to 2.5PN order in the above equation~(\ref{eq:stress-energytencsor_chk-2}). 
For that purpose, we compute the full stress-energy tensor up to the 2.5PN order. This involves 20 diagrams in the point particle sector (1 disconnected diagram, 17 tree-level diagrams and 2 one-loop diagrams) and 120 diagrams in the adiabatic tidal sector (6 tree-level diagrams, 49 one-loop diagrams and 65 two loop diagrams). Substituting this newly computed stress-energy components in the moment relation in equation~(\ref{eq:stress-energytencsor_chk-2}), we verify that the $\mathcal{T}^{00}$ component is consistent until 2PN order.
By applying these checks, we ensure that our stress-energy tensor components are consistent with the expected conservation laws, providing a robust validation up to the 2PN order.

\subsection{Coordinate transformations}

Having the freedom of making a coordinate transformation is very crucial in the post-Newtonian computation, so that one can change gauges in the Lagrangian and in the stress-energy tensor. 
In this section we describe the procedure to make a coordinates transformation. For this we start with the total effective action given in equation \eqref{eq_tot_eff_action}.
The action under a coordinate transformation $\textbf{x}_{(a)} \rightarrow \textbf{x}_{(a)} + \delta \textbf{x}_{(a)}$ changes
by
\begin{align}
    \delta ( \mathcal{L}_{\text{eff}} + \Gamma_{\text{eff}}[\bar{h}] ) = 
    \left( \frac{\delta \mathcal{L}_{\text{eff}}}{\delta \textbf{x}_{(a)}^i} + \frac{\delta \Gamma_{\text{eff}}[\bar{h}]}{\delta \textbf{x}_{(a)}^i} \right)~ \delta \textbf{x}_{(a)}^i + \mathcal{O}\left(\delta \textbf{x}_{(a)}^2\right) \, ,
\end{align}
where we consider the radiation graviton as background field.
We use this to remove the acceleration and its higher order time derivatives from the Lagrangian. When the 
equation of motion
(EOM)
is linear in $\textbf{a}_{(a)}$ at LO,
we can construct a perturbatively small $\delta \textbf{x}_{(a)}$ such that 
terms depending on $\textbf{a}_{(a)}$ drop out in the transformed effective Lagrangian. 
Similarly, for terms involving higher-order time derivatives of $\textbf{a}_{(a)}$,  one can take $\delta \textbf{x}_{(a)}$ to be a total time derivative such that the higher-order time derivatives cancel upon partial integration.
In general, the $\mathcal{O}\left(\delta \textbf{x}_{(a)}^2\right)$ contributions have to be kept, but will turn out to be negligible for the explicit steps outlined below, making the procedure equivalent to insertion of EOM~\cite{Damour:1990jh}, computed now from the total effective action. 
The removal of higher-order time derivatives through this process changes the gauge of the system, which also necessitates a consistent modification of the stress-energy tensor. This ensures that the resulting gauge remains consistent throughout the entire calculation.
Moreover, this procedure would make the conversion to EOB co-ordinates \cite{Gamboa:2024imd} seamless at the level of the action for non-trivial orbital configurations of the binary.

We apply the above procedure to eliminate higher-order time derivatives from the Lagrangian, resulting in a form that depends only on the positions $\textbf{x}_{(a)}$, velocities $\textbf{v}_{(a)}$, and the tidal parameter $\lambda_{(a)}$. The modified Lagrangian 
is presented in the appendix \ref{app_non_tidal} and \ref{app_tidal}, and 
also provided as an ancillary file \texttt{Lagrangian.m}. 
We also modify the stress-energy tensor by eliminating higher-order time derivatives by substituting the equation of motion computed by the modified Lagrangian. This is allowed since the stress-energy is on-shell. The resulting expressions only depends on $\textbf{x}_{(a)}$, velocities $\textbf{v}_{(a)}$ and are provided in the ancillary file \texttt{Stress\_Energy\_Tensor.m}.

\subsection{Mapping from stress-energy to multipoles}
\label{sec_sub_SE_to_Multipoles}

The effective action to describe the long distance gravitational degrees of freedom is given by worldline effective theory for the binary system treated as a point particle with multipole moments. This action for the multipole moments \footnote{In d-dimensions, there is one additional Weyl type multipole moment \cite{Amalberti:2023ohj} that is absent in three dimensions. This multipole moment plays a crucial role when the stress-energy tensor contains divergences \cite{Blanchet:2023bwj,Amalberti:2024jaa}. In our case, since the stree-energy tensor is finite, this extra multipole moment can be ignored.} of the binary is,
\begin{align}
    S=-\frac{1}{2} \int dt \Bigg(& \mathcal{M} \bar{h}_{00} + \textbf{G}^i \partial_i\bar{h}_{00} + 2 \textbf{P}^i \bar{h}_{0i} + \textbf{L}^{ij} \partial_i \bar{h}_{j0} \nonumber\\ 
    & +\sum_{l=2}^\infty \frac{1}{l!} \mathcal{I}^L \partial_{L-2} E_{i_i i_{l-1}} - \sum_{l=2}^\infty \frac{2l}{(l+1)!} \mathcal{J}^{i|L} \partial_{L-2} B_{i|i_i i_{l-1}} \Bigg)  \, ,
\end{align}
where, $\mathcal{M}$ is the mass, $\textbf{G}^i$ is the center-of-mass, $\textbf{P}^i$ is the linear momentum, $\textbf{L}^i$ is the angular momentum, $\mathcal{I}^L$ are mass multipoles and $\mathcal{J}^L = \epsilon^{i_l a b} \mathcal{J}_{d=3}^{a|bL-1} $  
are current multipoles. Here the $E_{(a)\mu\nu}= - c^2 R_{\mu\alpha\nu\beta}u_{(a)}^\alpha u_{(a)}^\beta/z_{(a)}^2$ is the electric component and $B_{(a)\alpha|\mu\nu}= c R_{\mu\alpha\nu\beta} u_{(a)}^\beta/z_{(a)}$ is the magnetic component of the Riemann tensor.

The matching of effective one-point function and the above action for the multipole moments of a binary can be achieved by decomposing the stress-energy tensor in terms of irreducible representations of SO(d), to obtain a relation between the stress-energy tensor and different multipole moments. This matching was done in Ref. \cite{Ross:2012fc} in three dimensions and for generic dimensions it is derived recently in Ref.~\cite{Amalberti:2023ohj}, which is given as follows:
    \begin{align}
        \mathcal{M} &= \int d^d\textbf{x} \mathcal{T}_{\text{eff}}^{00}  \, ;\\
        \textbf{G}^i &=\frac{1}{\mathcal{M}} \int d^d\textbf{x} \mathcal{T}_{\text{eff}}^{00} \textbf{x}^i  \, ;\\
        \textbf{P}^i &= \int d^d\textbf{x} \mathcal{T}_{\text{eff}}^{0i}  \, ;\\
        \textbf{L}^{ij} &= \int d^d\textbf{x} (\mathcal{T}_{\text{eff}}^{0i}\textbf{x}^j- \mathcal{T}_{\text{eff}}^{0j}\textbf{x}^i)  \, ;
    \end{align}
    \begin{align}
    \mathcal{I}^L
    = &  
    \sum_{p=0}^{\infty}\,\frac{\Gamma\left(\frac{d}{2}+ l\right)}{2^{2p}p!\,\Gamma\left(\frac{d}{2}+ l+p\right)}\left(1+\frac{4p\left(d-1\right)\left(d+ l+p-2\right)}{\left(d-2\right)\left(d+ l-1\right)\left(d+ l-2\right)}\right) \left[ \int  d^d\mathbf{x}\,\partial_t^{2p}\mathcal{T}_{\text{eff}}^{00}\hat{\textbf{x}}^{L} r^{2p} \right]_{\texttt{STF}} \nonumber\\
    &  
    -\sum_{p=0}^{\infty}\,\frac{\Gamma\left(\frac{d}{2}+ l\right)}{2^{2p}p!\,\Gamma\left(\frac{d}{2}+ l+p\right)}\frac{2\left(d-1\right)\left(d+ l+2p-1\right)}{\left(d-2\right)\left(d+ l-1\right)\left(d+ l-2\right)} \left[ \int  d^d\mathbf{x}\,\partial_t^{2p+1} \mathcal{T}_{\text{eff}}^{0a} \textbf{x}^a \hat{\textbf{x}}^{L} r^{2p}\right]_{\texttt{STF}}\nonumber\\
    & 
    +\sum_{p=0}^{\infty}\,\frac{\Gamma\left(\frac{d}{2}+ l\right)}{2^{2p}p!\,\Gamma\left(\frac{d}{2}+ l+p\right)}\frac{1}{(d-2)}\left(1+\frac{2p\left(d-1\right)}{\left(d+ l-1\right)\left(d+ l-2\right)}\right) \left[ \int  d^d\mathbf{x}\,\partial_t^{2p}\mathcal{T}_{\text{eff}}^{aa}\,\hat{\textbf{x}}^{L} r^{2p}\right]_{\texttt{STF}}\nonumber\\
    &
    +\sum_{\,p=0}^{\infty}\,\frac{\Gamma\left(\frac{d}{2}+ l\right)}{2^{2p}p!\,\Gamma\left(\frac{d}{2}+ l+p\right)}\frac{\left(d-1\right)}{\left(d-2\right)\left(d+ l-1\right)\left(d+ l-2\right)} \left[ \int  d^d\mathbf{x}\,\partial_t^{2p+2} \mathcal{T}_{\text{eff}}^{ab} x^{ab} \hat{\textbf{x}}^{L} r^{2p} \right]_{\texttt{STF}}   \, ;
    \end{align}
    \begin{align}
    \mathcal{J}^{a|L}
    = &
    \sum_{p=0}^{\infty}\,\frac{\Gamma\left(\frac{d}{2}+ l\right)}{2^{2p}p!\,\Gamma\left(\frac{d}{2}+ l+p\right)}\left(1+\frac{2p}{\left(d+ l-1\right)}\right)\left[\int  d^d\mathbf{x}\,\partial_t^{2p}\mathcal{T}_{\text{eff}}^{0a}\hat{\textbf{x}}^{L} r^{2p}\right]_{\substack{\texttt{STF-}L\\ [ai_l] }}\nonumber\\
    &%
    -
    \sum_{p=0}^{\infty}\,\frac{\Gamma\left(\frac{d}{2}+ l\right)}{2^{2p}p!\,\Gamma\left(\frac{d}{2}+ l+p\right)}\frac{1}{\left( d+ l-1\right)}\,\left[\int  d^d\mathbf{x}\,\partial_t^{2p+1}\mathcal{T}_{\text{eff}}^{ab} \textbf{x}^b \hat{\textbf{x}}^{L} r^{2p}\right]_{\substack{\texttt{STF-}L\\ [ai_l] }}  \, .
    \end{align}
    Here, $\texttt{STF-}L$ is symmetrization over the $L$ indices and $[ai_l]$ means to antisymmetrize $a$ and $i_l$.
In the current analysis the stress-energy tensor up to 2PN does not contain any poles, and thus the three dimensional mapping is sufficient. The d-dimensional mapping plays a crucial role at 3PN \cite{Amalberti:2024jaa} and the additional evanescent couplings play an important role at 4PN \cite{Blanchet:2023bwj,Amalberti:2023ohj}.

\section{Results}
\label{sec_results}

In this section, we present the results for the energy and angular momentum fluxes, as well as the gravitational mode amplitudes and phases up to 2PN order.
We express these results in terms of the following parameters of the binary system, including the total mass $M=m_{(1)}+m_{(2)}$, the reduced mass $\mu=m_{(1)}m_{(2)}/M$, the mass ratio $q=m_{(1)}/m_{(2)}$, the symmetric mass ratio $\nu=\mu/M$, and the antisymmetric mass ratio $\delta=(m_{(1)}-m_{(2)})/M$. These parameters are related by the following relation,
\begin{align}
\nu=\frac{m_{(1)} m_{(2)}}{M^2}=\frac{\mu}{M}=\frac{q}{(1+q)^2} = \frac{(1-\delta^2)}{4} \,.
\end{align}
Furthermore, we work in the center-of-mass frame, which we impose by setting $\textbf{G}^i=0$. This allows us to relate the general coordinates ($\textbf{x}_{(a)}$ and $\textbf{v}_{(a)}$) to the relative center-of-mass coordinates $\textbf{x} = \textbf{x}_{(1)}-\textbf{x}_{(2)}$ and velocity $\textbf{v}=\textbf{v}_{(1)}-\textbf{v}_{(2)}$.
To simplify the expressions and make them more compact, we introduce dimensionless variables as follows:
\begin{align}
&\widetilde{\textbf{v}}=\frac{\textbf{v}}{c} \, ,\quad
\widetilde{\bm{r}}=\frac{\bm{r}}{G_N M/c^2} \, ,\quad
\widetilde{\mathcal{L}}= \frac{\mathcal{L}}{\mu c^2} \, , \quad \widetilde{\lambda}= \frac{\lambda}{G_N^4M^5/c^{10}} \, , \\
\label{eq_rescale}
\widetilde{\mathcal{M}}= &\frac{\mathcal{M}}{\mu} ~,\quad  \widetilde{\textbf{L}}= \frac{\textbf{L} }{\mu G M/c  } ~,\quad \widetilde{\mathcal{I}}^L= \frac{\mathcal{I}^L}{\mu (G M / c^2 )^l } ~,\quad \widetilde{\mathcal{J}}^L= \frac{\mathcal{J}^L}{\mu (G M/ c^2)^l } \, .
\end{align}
We also introduce a symmetric and anti-symmetric combination of the Love numbers, defined using
\begin{align}
    \widetilde{\lambda}_{(\pm)} &=
\frac{m_{(2)}}{m_{(1)}} \widetilde{\lambda}_{(1)}
\,\pm\,
\frac{m_{(1)}}{m_{(2)}} \widetilde{\lambda}_{(2)} \,,
\end{align}
which simplifies to $\widetilde{\lambda}_{(+)} = \widetilde{\lambda}_{(1)}=\widetilde{\lambda}_{(2)}$ and $\widetilde{\lambda}_{(-)}=0$, when the two objects in the binary are identical.
For circular orbits, we use the PN parameter $x=(GM \omega /c^3)^{2/3}$. 
We begin by presenting the multipole moments of the system. Using these moments, we then compute the energy flux and angular momentum flux for circular orbits. Finally, applying the energy balance relations, we derive the modes of the emitted gravitational waves and present their amplitude and phase.

\subsection{Multipole moments}

We compute multipole moments of the binary from the effective one-point function on the center-of-mass coordinatess using the algorithm given section \ref{sec_com_algo}.
Using this procedure the mass of the binary is given by,
\begin{align}
    \widetilde{\mathcal{M}} = \widetilde{\mathcal{M}}^{\rm pp} + \widetilde{\mathcal{M}}^{\rm AT}  \, ,
\end{align}
where,
\begin{align}
    \widetilde{\mathcal{M}}^{\rm pp} &= \frac{1}{\nu} + \bigg\{ \frac{\widetilde{v}^2}{2}-\frac{1}{\widetilde{r}} \bigg\} + \bigg\{ \left(\frac{3}{8}-\frac{9 \nu }{8}\right) \widetilde{v}^4 + \frac{1}{\widetilde{r}} \left[\frac{\nu  (\textbf{n}\cdot\widetilde{\textbf{v}})^2}{2}+\left(\frac{\nu }{2}+\frac{3}{2}\right) \widetilde{v}^2\right]+\frac{1}{2\widetilde{r}^2} \bigg\} + \mathcal{O}\left(\frac{1}{c^5}\right)  \, ;\\
    \widetilde{\mathcal{M}}^{\rm AT} &=   \frac{\widetilde{\lambda}_{(1)}}{\widetilde{r}^5}\bigg\{ -\frac{3}{2 q \widetilde{r}} \bigg\}+ \frac{\widetilde{\lambda}_{(1)}}{\widetilde{r}^5}\bigg\{ \frac{1}{\widetilde{r}} \left(\frac{\nu  \widetilde{v}^2}{4}-\frac{3 \nu  (\textbf{n}\cdot\widetilde{\textbf{v}})^2}{2}\right)-\frac{7 \nu  }{2\widetilde{r}^2} \nonumber\\
    &+ \frac{1}{q}\bigg[ \frac{1}{\widetilde{r}} \left(\left(3 \nu -\frac{9}{2}\right) (\textbf{n}\cdot\widetilde{\textbf{v}})^2+\left(\nu +\frac{15}{4}\right) \widetilde{v}^2\right)+\left(\frac{21}{2}-\frac{7 \nu }{2}\right) \frac{1}{\widetilde{r}^2} \bigg]  \bigg\}+ (1\leftrightarrow 2) + \mathcal{O}\left(\frac{1}{c^5}\right)
\end{align}
Computing the above equation on circular orbits exactly gives the result of equation (5.15) of Ref. \cite{Mandal:2023lgy} which also serves as a consistency check of our computation.
The linear momentum is given by $\textbf{P}^{i} =  \dot{\textbf{G}}^{i} =0$ using the conservation of the stress-energy tensor. The angular momentum vector is $\textbf{L}^{i}= \epsilon^{iab} \textbf{L}^{ab}$ and is given by,
\begin{align}
    \widetilde{\textbf{L}}^{i} = \Big(\widetilde{\textbf{L}}^{\rm pp}\Big)^{i} + \Big(\widetilde{\textbf{L}}^{\rm AT}\Big)^{i}  \, ,
\end{align}
where,
\begin{align}
    \Big(\widetilde{\textbf{L}}^{\rm pp}\Big)^{i}=& \Big[ \epsilon^{iab }\textbf{n}^a\widetilde{\textbf{v}}^b \Big] \bigg\{ \widetilde{r} + \bigg[ 3+ \nu +\left(\frac{1}{2}-\frac{3 \nu }{2}\right) \widetilde{r} \widetilde{v}^2 \bigg]\bigg\} + \mathcal{O}\left(\frac{1}{c^5}\right)  \, ; \\
   \Big(\widetilde{\textbf{L}}^{\rm AT}\Big)^{i}=&\Big[ \epsilon^{iab }\textbf{n}^a\widetilde{\textbf{v}}^b \Big] \frac{\widetilde{\lambda}_{(1)}}{\widetilde{r}^5} \bigg\{ \frac{\nu }{2}+\left(2 \nu +\frac{15}{2}\right) \frac{1}{q} \bigg\}  +(1\leftrightarrow 2) + \mathcal{O}\left(\frac{1}{c^5}\right)  \, .
\end{align}
All the above given quantities are non-radiative since they are conserved due to the conservation of the stress-energy tensor. The dynamic quantities that contribute to the gravitational radiation are given in the next sections and their expressions are provided in an ancillary file \texttt{Multipole_Moments.m}

\subsubsection{Mass multipoles}
The key mass moment required for the computation of the fluxes, phase, and the dominant quadrupolar mode amplitude $(2,\pm 2)$ of the emitted waves is the mass quadrupole moment, which must be determined to the highest PN precision. Here, we present the mass quadrupole moment, denoted as $\mathcal{I}^{ij}$, up to 2PN order, along with the mass octupole $\mathcal{I}^{ijk}$ up to 1PN order, and the mass decapole $\mathcal{I}^{ijkl}$ at 0PN order. These moments are essential for obtaining the fluxes and gravitational wave modes~\footnote{In this article, we only present the dominant quadrupolar mode (\(l=2, m=2\)). For this purpose, higher-order multipoles are required but only up to a lower PN precision compared to the mass quadrupole.} up to 2PN order.

We begin by decomposing the moments in the point-particle sector and the adiabatic sector as follows:
\begin{align}
    \widetilde{\mathcal{I}}^{L} = \Big(\widetilde{\mathcal{I}}^{\rm pp}\Big)^{L} + \Big(\widetilde{\mathcal{I}}^{\rm AT}\Big)^{L},
\end{align}
where the mass moments in the point-particle sector $\left(\widetilde{\mathcal{I}}^{\rm pp}\right)^{L}$ are provided in Appendix \ref{app_pp_multimom}, and the mass moments in the adiabatic tidal sector are presented below. We present the contributions for different mass multipole moments at different PN orders as follows:
\begin{align}
    \Big(\widetilde{\mathcal{I}}^{\rm AT}\Big)^{ij} =& \Big(\widetilde{\mathcal{I}}_{\rm 0PN}^{\rm AT}\Big)^{ij}+\Big(\widetilde{\mathcal{I}}_{\rm 1PN}^{\rm AT}\Big)^{ij}+\Big(\widetilde{\mathcal{I}}_{\rm 2PN}^{\rm AT}\Big)^{ij} + \mathcal{O}\left(\frac{1}{c^5}\right) \, ; \\
    \Big(\widetilde{\mathcal{I}}^{\rm AT}\Big)^{ijk} =& \Big(\widetilde{\mathcal{I}}_{\rm 0PN}^{\rm AT}\Big)^{ijk}+\Big(\widetilde{\mathcal{I}}_{\rm 1PN}^{\rm AT}\Big)^{ijk} + \mathcal{O}\left(\frac{1}{c^3}\right) \, ; \\
    \Big(\widetilde{\mathcal{I}}^{\rm AT}\Big)^{ijkl} =& \Big(\widetilde{\mathcal{I}}_{\rm 0PN}^{\rm AT}\Big)^{ijkl} + \mathcal{O}\left(\frac{1}{c}\right) \, ,
\end{align}
where the individual contributions to different PN orders are,
\begin{align}
    \Big(\widetilde{\mathcal{I}}_{\rm 0PN}^{\rm AT}\Big)^{ij} = & \Big[ \textbf{n}^i\textbf{n}^j \Big]_{ \texttt{\tiny STF}} \frac{\widetilde{\lambda}_{(1)}}{\widetilde{r}^5} \bigg\{ 3\left(1+\frac{1}{q}\right)\widetilde{r}^2 \bigg\}  +(1\leftrightarrow 2) \, ;
\end{align}
\begin{align}
    \Big(\widetilde{\mathcal{I}}_{\rm 1PN}^{\rm AT}\Big)^{ij} = &  
    \Big[ \textbf{n}^i\textbf{n}^j \Big]_{ \texttt{\tiny STF}} \frac{\widetilde{\lambda}_{(1)}}{\widetilde{r}^5} \bigg\{ \widetilde{r}^2 \left(\left(-\frac{40 \nu }{7}-\frac{15}{2}\right) (\textbf{n}\cdot\widetilde{\textbf{v}})^2+\left(6-\frac{13 \nu }{7}\right) \widetilde{v}^2\right)+\left(\frac{2 \nu }{7}-\frac{15}{2}\right) \widetilde{r} \nonumber\\
    & \quad\quad\quad\quad + \frac{1}{q} \left[\widetilde{r}^2 \left(\left(\frac{185}{14}-\frac{40 \nu }{7}\right) (\textbf{n}\cdot\widetilde{\textbf{v}})^2+\left(\frac{55}{7}-\frac{13 \nu }{7}\right) \widetilde{v}^2\right)+\left(\frac{43 \nu }{14}-\frac{75}{7}\right) \widetilde{r}\right] \bigg\} \nonumber\\ 
    & + \Big[ \textbf{n}^i\widetilde{\textbf{v}}^j \Big]_{ \texttt{\tiny STF}}  \frac{\widetilde{\lambda}_{(1)}}{\widetilde{r}^5} \bigg\{ \left(\frac{130 \nu }{7}-\frac{214}{7}\right) (\textbf{n}\cdot\widetilde{\textbf{v}}) q \widetilde{r}^2+\left(\frac{130 \nu }{7}-6\right) (\textbf{n}\cdot\widetilde{\textbf{v}}) \widetilde{r}^2 \bigg\} \nonumber \\
    & + \Big[ \widetilde{\textbf{v}}^i\widetilde{\textbf{v}}^j \Big]_{ \texttt{\tiny STF}} \frac{\widetilde{\lambda}_{(1)}}{\widetilde{r}^5} \bigg\{\left(\frac{59}{7}-\frac{38 \nu }{7}\right) q \widetilde{r}^2+\left(3-\frac{38 \nu }{7}\right) \widetilde{r}^2 \bigg\}  +(1\leftrightarrow 2) \, ;
\end{align}
\begin{align}
    \Big(\widetilde{\mathcal{I}}_{\rm 2PN}^{\rm AT}\Big)^{ij} = & 
    \Big[ \textbf{n}^i\textbf{n}^j \Big]_{ \texttt{\tiny STF}} \frac{\widetilde{\lambda}_{(1)}}{\widetilde{r}^5} \bigg\{ \left(\frac{617 \nu ^2}{84}+\frac{4237 \nu }{84}+\frac{285}{28}\right) + \widetilde{r} \bigg(\left(-\frac{19 \nu ^2}{8}+\frac{361 \nu }{4}-\frac{639}{8}\right) (\textbf{n}\cdot\widetilde{\textbf{v}})^2 \nonumber\\
    &\quad\quad\quad\quad +\left(-\frac{605 \nu ^2}{168}-\frac{197 \nu }{12}+\frac{139}{8}\right) \widetilde{v}^2\bigg) +\widetilde{r}^2 \bigg(\left(-15 \nu ^2-85 \nu +\frac{105}{8}\right) (\textbf{n}\cdot\widetilde{\textbf{v}})^4 \nonumber\\
    &\quad\quad\quad\quad +\left(\frac{205 \nu ^2}{14}+15 \nu -\frac{45}{2}\right) (\textbf{n}\cdot\widetilde{\textbf{v}})^2 \widetilde{v}^2+\left(\frac{61 \nu ^2}{14}-\frac{96 \nu }{7}+6\right) \widetilde{v}^4\bigg) \nonumber\\
    & \quad\quad\quad\quad + \frac{1}{q} \bigg[ \left( \frac{296 \nu ^2}{21}+\frac{453 \nu }{14}+\frac{91}{6} \right) + \widetilde{r} \bigg(\left(-\frac{985 \nu ^2}{56}+\frac{4639 \nu }{28}+\frac{2187}{56}\right) (\textbf{n}\cdot\widetilde{\textbf{v}})^2 \nonumber\\
    &\quad\quad\quad\quad +\left(-\frac{837 \nu ^2}{56}-\frac{772 \nu }{21}+\frac{533}{168}\right) \widetilde{v}^2\bigg) + \widetilde{r}^2 \bigg(\left(-15 \nu ^2-120 \nu +\frac{365}{8}\right) (\textbf{n}\cdot\widetilde{\textbf{v}})^4\nonumber\\
    &\quad\quad\quad\quad +\left(\frac{205 \nu ^2}{14}-\frac{75 \nu }{7}+30\right) (\textbf{n}\cdot\widetilde{\textbf{v}})^2 \widetilde{v}^2+\left(\frac{61 \nu ^2}{14}-\frac{123 \nu }{7}+\frac{54}{7}\right) \widetilde{v}^4\bigg)\bigg]
    \bigg\}  \nonumber\\
    & + \Big[ \textbf{n}^i\widetilde{\textbf{v}}^j \Big]_{ \texttt{\tiny STF}} \frac{\widetilde{\lambda}_{(1)}}{\widetilde{r}^5} \bigg\{ \widetilde{r}^2 \left(\left(\frac{110 \nu ^2}{7}+\frac{1055 \nu }{7}+15\right) (\textbf{n}\cdot\widetilde{\textbf{v}})^3+\left(-\frac{286 \nu ^2}{7}+\frac{27 \nu }{7}-6\right) (\textbf{n}\cdot\widetilde{\textbf{v}}) \widetilde{v}^2\right) \nonumber\\
    &\quad\quad\quad\quad+\frac{1}{q} \bigg[ \widetilde{r}^2 \bigg(\left(\frac{110 \nu ^2}{7}+\frac{1905 \nu }{7}-\frac{1055}{7}\right) (\textbf{n}\cdot\widetilde{\textbf{v}})^3+\left(-\frac{286 \nu ^2}{7}-\frac{3 \nu }{7}-\frac{48}{7}\right) (\textbf{n}\cdot\widetilde{\textbf{v}}) \widetilde{v}^2\bigg) \nonumber\\
    &\quad\quad\quad\quad+\left(\frac{1909 \nu ^2}{42}-\frac{645 \nu }{7}-\frac{1931}{42}\right) (\textbf{n}\cdot\widetilde{\textbf{v}}) \widetilde{r}\bigg] \bigg\}  \nonumber\\
    & + \Big[ \widetilde{\textbf{v}}^i\widetilde{\textbf{v}}^j \Big]_{ \texttt{\tiny STF}} \frac{\widetilde{\lambda}_{(1)}}{\widetilde{r}^5} \bigg\{ \left(\frac{25 \nu ^2}{63}-\frac{25 \nu }{63}+\frac{5}{63}\right) (\textbf{n}\cdot\widetilde{\textbf{v}})^2 \widetilde{r}^2+\left(\frac{733 \nu ^2}{126}-\frac{337 \nu }{126}+\frac{41}{126}\right) \widetilde{r}^2 \widetilde{v}^2 \nonumber\\
    & \quad\quad\quad\quad +\left(-\frac{985 \nu ^2}{189}-\frac{335 \nu }{189}+\frac{106}{27}\right) \widetilde{r} \bigg\}  +(1\leftrightarrow 2) \, ;
\end{align}
\begin{align}
    \Big(\widetilde{\mathcal{I}}_{\rm 0PN}^{\rm AT}\Big)^{ijk} =& \Big[ \textbf{n}^i\textbf{n}^j\textbf{n}^k \Big]_{ \texttt{\tiny STF}} \frac{\widetilde{\lambda}_{(1)}}{\widetilde{r}^5} \bigg\{ -9 \frac{1}{q} \widetilde{r}^3 \bigg\} +(1\leftrightarrow 2) \, ; \\
    \Big(\widetilde{\mathcal{I}}_{\rm 1PN}^{\rm AT}\Big)^{ijk} =& \Big[ \textbf{n}^i\textbf{n}^j\textbf{n}^k \Big]_{ \texttt{\tiny STF}} \frac{\widetilde{\lambda}_{(1)}}{\widetilde{r}^5} \bigg\{ \widetilde{r}^3 \left(\frac{105 \nu  (\textbf{n}\cdot\widetilde{\textbf{v}})^2}{2}+3 \nu  \widetilde{v}^2\right)+\left(9 \nu ^2-\frac{9 \nu }{2}\right) \widetilde{r}^2 \nonumber\\
    &\quad\quad\quad\quad\quad+ \frac{1}{q} \left(\widetilde{r}^3 \left((60 \nu -30) (\textbf{n}\cdot\widetilde{\textbf{v}})^2+\left(\frac{39 \nu }{2}-\frac{51}{2}\right) \widetilde{v}^2\right)+\left(9 \nu ^2-\frac{39 \nu }{2}+42\right) \widetilde{r}^2\right) \bigg\}    \nonumber\\
    &+ \Big[ \textbf{n}^i\textbf{n}^j\widetilde{\textbf{v}}^k \Big]_{ \texttt{\tiny STF}} \frac{\widetilde{\lambda}_{(1)}}{\widetilde{r}^5} \bigg\{ (90-126 \nu ) (\textbf{n}\cdot\widetilde{\textbf{v}}) \frac{1}{q} \widetilde{r}^3-72 \nu  (\textbf{n}\cdot\widetilde{\textbf{v}}) \widetilde{r}^3 \bigg\} \nonumber\\
    &+ \Big[ \textbf{n}^i\widetilde{\textbf{v}}^j\widetilde{\textbf{v}}^k \Big]_{ \texttt{\tiny STF}} \frac{\widetilde{\lambda}_{(1)}}{\widetilde{r}^5} \bigg\{ (42 \nu -30) \frac{1}{q} \widetilde{r}^3+21 \nu  \widetilde{r}^3 \bigg\} +(1\leftrightarrow 2) \, ;
\end{align}
\begin{align}
    \Big(\widetilde{\mathcal{I}}_{\rm 0PN}^{\rm AT}\Big)^{ijkl} =& \Big[ \textbf{n}^i\textbf{n}^j\textbf{n}^k\textbf{n}^l \Big]_{ \texttt{\tiny STF}} \frac{\widetilde{\lambda}_{(1)}}{\widetilde{r}^5} \bigg\{ (18-18 \nu ) \frac{1}{q} \widetilde{r}^4-18 \nu  \widetilde{r}^4 \bigg\} +(1\leftrightarrow 2) \, .
\end{align}

\subsubsection{Current multipoles}

For the computation of fluxes, phase, and the dominant quadrupolar mode amplitudes $(2,\pm 2)$ of the emitted waves up to 2PN order, we require the current quadrupole moment $\mathcal{J}^{ij}$ up to 1PN order and the current octupole moment $\mathcal{J}^{ijk}$ at 0PN order. These are presented in this section. We first decompose the current multipoles in the point particle sector and the adiabatic sector as follows:
\begin{align}
    \widetilde{\mathcal{J}}^{L} = \Big(\widetilde{\mathcal{J}}^{\rm pp}\Big)^{L} + \Big(\widetilde{\mathcal{J}}^{\rm AT}\Big)^{L},
\end{align}
where the current multipoles in the point-particle sector $\left(\widetilde{\mathcal{J}}^{\rm pp}\right)^{L}$ are provided in Appendix \ref{app_pp_multimom}, and in the adiabatic tidal sector are presented below. 
We present the contributions for different current multipole moments at different PN orders as follows:
\begin{align}
    \Big(\widetilde{\mathcal{J}}^{\rm AT}\Big)^{ij} &= \Big(\widetilde{\mathcal{J}}_{\rm 0PN}^{\rm AT}\Big)^{ij} + \Big(\widetilde{\mathcal{J}}_{\rm 1PN}^{\rm AT}\Big)^{ij} + \mathcal{O}\left(\frac{1}{c^3}\right)  \, ; \\
    \Big(\widetilde{\mathcal{J}}^{\rm AT}\Big)^{ijk} &= \Big(\widetilde{\mathcal{J}}_{\rm 0PN}^{\rm AT}\Big)^{ijk} + \mathcal{O}\left(\frac{1}{c}\right),
\end{align}
and the individual contributions at different PN orders are,
\begin{align}
    \Big(\widetilde{\mathcal{J}}_{\rm 0PN}^{\rm AT}\Big)^{ij} =& \Big[ \epsilon^{imn}\textbf{n}^j\textbf{n}^m\widetilde{\textbf{v}}^n \Big]_{ \texttt{\tiny STF}} \frac{\widetilde{\lambda}_{(1)}}{\widetilde{r}^5} \bigg\{  \frac{9}{2q} \widetilde{r}^2  \bigg\} +(1\leftrightarrow 2) \,  ; \\
    \Big(\widetilde{\mathcal{J}}_{\rm 1PN}^{\rm AT}\Big)^{ij} =& \Big[ \epsilon^{imn}\textbf{n}^j\textbf{n}^m\widetilde{\textbf{v}}^n \Big]_{ \texttt{\tiny STF}} \frac{\widetilde{\lambda}_{(1)}}{\widetilde{r}^5} \bigg\{\widetilde{r}^2 \left(\frac{129 \nu  \widetilde{v}^2}{28}-\frac{240 \nu  (\textbf{n}\cdot\widetilde{\textbf{v}})^2}{7}\right)+\left(-\frac{60 \nu ^2}{7}-\frac{165 \nu }{14}\right) \widetilde{r} \nonumber\\
    &\quad\quad\quad\quad +\frac{1}{q} \left(\widetilde{r}^2 \left(\left(\frac{855}{28}-\frac{270 \nu }{7}\right) (\textbf{n}\cdot\widetilde{\textbf{v}})^2+\left(\frac{27 \nu }{28}+\frac{36}{7}\right) \widetilde{v}^2\right)+\left(-\frac{60 \nu ^2}{7}-\frac{18 \nu }{7}-\frac{57}{14}\right) \widetilde{r}\right)\bigg\} \nonumber\\
    &+ \Big[ \epsilon^{imn}\widetilde{\textbf{v}}^j\textbf{n}^m\widetilde{\textbf{v}}^n \Big]_{ \texttt{\tiny STF}} \frac{\widetilde{\lambda}_{(1)}}{\widetilde{r}^5} \bigg\{ \left(\frac{171 \nu }{7}-\frac{180}{7}\right) (\textbf{n}\cdot\widetilde{\textbf{v}}) \frac{1}{q} \widetilde{r}^2+\frac{213}{14} \nu  (\textbf{n}\cdot\widetilde{\textbf{v}}) \widetilde{r}^2 \bigg\}  +(1\leftrightarrow 2)  \, ;
\end{align}
\begin{align}
    \Big(\widetilde{\mathcal{J}}_{\rm 0PN}^{\rm AT}\Big)^{ijk} =& \Big[ \epsilon^{imn}\textbf{n}^j\textbf{n}^k\textbf{n}^m\widetilde{\textbf{v}}^n \Big]_{ \texttt{\tiny STF}} \frac{\widetilde{\lambda}_{(1)}}{\widetilde{r}^5} \bigg\{ (12-12 \nu ) \frac{1}{q} \widetilde{r}^3-12 \nu  \widetilde{r}^3 \bigg\} +(1\leftrightarrow 2) \, .
\end{align}

\subsection{Energy and angular momentum fluxes}
The total energy flux ($\mathcal{F}_E$) can be decomposed in terms of the instantaneous part ($\mathcal{F}_E|_{\rm inst.}$) and the tail contributions ($\mathcal{F}_E|_{\rm tails}$) to it.
The instantaneous part of the total energy flux are expressed in terms of the multipole moments of the binary as \cite{RevModPhys.52.299},
\begin{align}
    \mathcal{F}_E\Big|_{\rm inst.}= \sum_{l=2}^{\infty} \Bigg[&\frac{G(l+1)(l+2)}{l(l-1)l!(2l+1)!!}\Bigg<\frac{d^{l+1}\bigl(\mathcal{I}^L\bigr)}{dt^{l+1}} \frac{d^{l+1} \bigl(\mathcal{I}^L\bigr)}{dt^{l+1}}\Bigg> \nonumber\\
    &+  \frac{4Gl(l+2)}{(l-1)(l+1)!(2l+1)!!}\Bigg<\frac{d^{l+1} \bigl(\mathcal{J}^L\bigr)}{dt^{l+1}}\frac{d^{l+1} \bigl(\mathcal{J}^L\bigr)}{dt^{l+1}}\Bigg> \Bigg]  \, .
\end{align}
By substituting the computed multipole moments from the previous section, we obtain the result for the instantaneous part of the energy flux up to 2PN order for generic orbits in center-of-mass coordinates. The detailed expression is provided in the ancillary file 	\texttt{Fluxes.m}.

The tail contribution to the flux arises due to the interaction of the emitted gravitational radiation from the mass multipoles with the curved background spacetime generated by the mass monopole of the binary. This tail term represents the back-scattering of radiation off the curved geometry, effectively modifying the radiation observed at infinity. As we are interested in obtaining expressions for the gauge-invariant observables on a circular orbit, we present the results of the energy flux in the circular orbit.

The computation of the tail contribution for circular orbits is detailed in~\cite{Goldberger:2009qd}, where equation (110) expresses it in terms of the leading-order contribution to the instantaneous flux and the PN parameter $x$ for circular orbits as,
\begin{align}
    \mathcal{F}_E\Big|_{\rm tails} = 4 \pi x^{3/2} 
     \mathcal{F}_E^{\rm LO}|_{\rm inst.} + \mathcal{O}(x^{5/2}) \, .
\end{align}
We substitute the relations for the circular orbit, as given in \eqref{eq_cir_orb_pp} and \eqref{eq_cir_orb_AT}, in the expression of the instantaneous energy flux and combine the tail contributions with it to obtain the total energy flux up to 2PN for circular orbits in center-of-mass coordinates. Now, we decompose this energy flux in the point particle sector and the adiabatic sector as follows:
\begin{align}
\label{eq_flux}
    \mathcal{F}_E=& \mathcal{F}_E^{\rm pp}+ \mathcal{F}_E^{\rm AT} \, ,
\end{align}
where the energy flux in the point particle sector up to 2PN order~\cite{Blanchet:1995fg,Blanchet:1995ez,Will:1996zj,Blanchet:1996wx} is provided by 
\begin{align}
    \mathcal{F}_E^{\rm pp}=& \bigg\{\frac{32 \nu^2}{5}\bigg\} x^5 + \bigg\{ -\frac{56 \nu ^3}{3}-\frac{2494 \nu ^2}{105} \bigg\} x^6 +  4 \pi \bigg\{\frac{32 \nu^2}{5}\bigg\}  x^{3/2} \nonumber\\
    &+ \bigg\{ \frac{208 \nu ^4}{9}+\frac{37084 \nu ^3}{315}-\frac{89422 \nu ^2}{2835} \bigg\} x^7 + \mathcal{O}\left(x^{15/2}\right) \, ,
\end{align}
and the adiabatic tidal contribution up to 2PN order is given by,
\begin{align}
    \mathcal{F}_E^{\rm AT}=&\bigg\{  \left(\frac{768 \nu ^2}{5}+\frac{192 \nu }{5}\right) \widetilde{\lambda}_{(+)} +  \frac{192}{5} \delta\widetilde{\lambda}_{(-)} \bigg\} x^{10} \nonumber\\
    & + \bigg\{ \left(-992 \nu ^3-\frac{9736 \nu ^2}{35}-\frac{1408 \nu }{35}\right) \widetilde{\lambda}_{(+)}+ \left( -\frac{184 \nu ^2}{5}-\frac{1408 \nu }{35} \right) \delta\widetilde{\lambda}_{(-)} \bigg\} x^{11} \nonumber\\
    &+4\pi \bigg\{  \left(\frac{768 \nu ^2}{5}+\frac{192 \nu }{5}\right) \widetilde{\lambda}_{(+)} +  \frac{192}{5} \delta\widetilde{\lambda}_{(-)} \bigg\} x^{23/2} \nonumber\\
    & + \bigg\{ \left(3088 \nu ^4+\frac{63692 \nu ^3}{35}-\frac{1299706 \nu ^2}{945}+\frac{5344 \nu }{45}\right) \widetilde{\lambda}_{(+)} \nonumber\\
    & \quad\quad\quad+   \left( -\frac{11116 \nu ^3}{15}+\frac{149566 \nu ^2}{105}+\frac{5344 \nu }{45} \right) \delta\widetilde{\lambda}_{(-)} \bigg\} x^{12} + \mathcal{O}\left(x^{25/2}\right) \, .
\end{align}
The adiabatic tidal contribution to the flux $\mathcal{F}_E^{\rm AT}$ up to 2PN order represents the main result of this work. 
Upon comparison, the authors of Ref.~\cite{Henry:2020ski} confirmed our result of $\mathcal{F}_E^{\rm AT}$ up to 2PN order in Ref.~\cite{Henry:2020skiErr}.

On the other hand, the instantaneous contribution to the angular momentum flux ($\mathcal{F}_J|_{\rm inst.}$) can be written in terms of the multipole moments as \cite{RevModPhys.52.299},
\begin{align}
    (\mathcal{F}_J)^i\Big|_{\rm inst.}= \epsilon^{iab} \sum_{l=2}^{\infty}  \Bigg[&\frac{G(l+1)(l+2)}{(l-1)l!(2l+1)!!}\Bigg< \frac{d^{l} \bigl(\mathcal{I}^{aL-1}\bigr) }{dt^{l}} \frac{d^{l+1} \bigl(\mathcal{I}^{bL-1}\bigr)}{dt^{l+1}} \Bigg>\nonumber\\
    &+ \frac{4Gl^2(l+2)}{(l-1)(l+1)!(2l+1)!!}\Bigg< \frac{d^{l} \bigl(\mathcal{J}^{aL-1} \bigr)}{dt^{l}}\frac{d^{l+1} \bigl(\mathcal{J}^{bL-1} \bigr)}{dt^{l+1}}\Bigg>\Bigg]   \, .
\end{align}
Using the multipole moments, computed in the previous section, we obtain the result of the instantaneous contribution to the angular momentum flux up to 2PN order for generic orbits in center-of-mass coordinates. We provide the detailed expressions in the \texttt{Fluxes.m} ancillary file. 

In the specialized case of circular orbits, using `the first law of binary mechanics' \cite{Friedman:2001pf,LeTiec:2011ab} the total angular momentum flux can be written in terms of the total energy flux using the relation,
\begin{align}
    \mathcal{F}_J^i= x^{3/2}  ~\Big(\epsilon^{ijk} \textbf{n}^j\hat{\textbf{v}}^k\Big)~ \mathcal{F}_E \, ,
\end{align}
where $\mathcal{F}_E$ is given in equation \eqref{eq_flux}. 
Using this relation for the tail part gives us the tail contribution to the angular momentum flux as $(\mathcal{F}_J)^i |_{\rm tails} = 4 \pi x^{3/2} (\mathcal{F}_J^{\rm LO})^i|_{\rm inst.} + \mathcal{O}(x^{5/2})$ \cite{Arun:2009mc}. Similarly, using this for the instantaneous part provides us a stringent check on our computations of the instantaneous angular momentum flux ($\mathcal{F}_J|_{\rm inst.}$).

\subsection{Modes of the waveform for circular orbits}
In this section, we present the dominant quadrupolar mode amplitudes $(2,\pm 2)$, which, along with the phase discussed in the next section, characterize the tidal effects on the emitted gravitational waveform.

We begin by decomposing the gravitational wave polarizations in terms of modes $\bar{h}_{lm}$ using spin-weighted spherical harmonics $\mathcal{Y}^{lm}_{-2}$ \footnote{We first choose a cartesian orthonormal triad $(\hat{\textbf{i}},\hat{\textbf{j}},\hat{\textbf{k}})$ such that the direction of GW propagation that we are interested in, is $\hat{\textbf{N}} = \hat{\textbf{i}} \sin \Theta \cos \Phi + \hat{\textbf{j}} \sin \Theta \sin \Phi + \hat{\textbf{k}} \cos \Theta $. 
The conversion of a \texttt{STF} tensor to spherical harmonic components is then obtained using
\begin{align}
    \alpha^L_{lm} = l! \sqrt{\frac{4\pi}{(2l+2)}\frac{2^m}{(l+m)!(l-m)!}} \Big[ \hat{\textbf{m}}^*_{M}\hat{\textbf{k}}_{L-M} \Big]_{\texttt{STF}} \, ,
\end{align}
where, the plane of the binary is encoded in $\hat{\textbf{m}} = 1/\sqrt{2} (\hat{\textbf{i}}+\ii\hat{\textbf{j}})$. For more details see Ref.~\cite{RevModPhys.52.299,Kidder:2007rt,Blanchet:2013haa}
}  of weight -2 given as \cite{Buonanno:2006ui,Kidder:2007rt}
\begin{align}
    \bar{h}_{+}-i\bar{h}_{-}=\sum_{l=2}^{\infty}\sum_{m=-l}^{l}\bar{h}_{lm}\mathcal{Y}^{lm}_{-2} (\Theta,\Phi) \, .
\end{align}
There are both instantaneous and tail contribution to the spherical harmonic modes.
The instantaneous contribution to the modes is computed using  
\cite{RevModPhys.52.299,Porto:2012as}
\begin{align}
    \bar{h}_{lm}\Big|_{\rm inst.}&=-\frac{G}{\sqrt{2}R c^{l+2}}\left( \frac{4}{l!}\sqrt{\frac{(l+1)(l+2)}{2l(l+1)}} \alpha^L_{lm} \frac{d^l \mathcal{I}^L}{dt^l} +\frac{\ii}{c} \frac{8}{l!}\sqrt{\frac{l(l+2)}{2(l+1)(l-1)}} \alpha^L_{lm}  \frac{d^l\mathcal{J}^L}{dt^l}\right)   \, ;\\
%
%
    \bar{h}_{lm}\Big|_{\rm tails}&=-\ii\frac{G^2 M \pi \sqrt{6}}{R c^{7}}  \alpha^{ij}_{lm} \frac{d^3 \mathcal{I}^{ij}}{dt^3}  +  \mathcal{O}\left( \frac{1}{c^8} \right) \, .
\end{align}

The spherical harmonic modes can be written in terms of the mode amplitudes\footnote{
Here we give results of $m\geq 0$. Using the symmetiries of the spherical harmonic functions, $\mathcal{H}_{l,-m} = (-1)^l \bar{\mathcal{H}}_{l,m}$.
}
 $\mathcal{H}_{lm}$ and an overall half-phase $\phi$ \footnote{
 The overall half-phase of the modes 
 gets modified at 4PN by the tail contribution given by \cite{Blanchet:1993ec,Porto:2012as} 
\begin{align}
    \bar{\phi} = \phi - \frac{2 G M \omega}{c^3}\log \left(\frac{\omega}{\omega_0}\right) \, ,
\end{align}
where the $\omega_0$ has to be fixed to a reference frequency corresponding to the detectors, for example the frequency at which the binary enters the detector band.
 }(given by the phase of the binary computed in section \ref{sec_phase}), can be written as
\begin{align}
    \bar{h}_{lm} = \frac{1}{R} \frac{GM}{c^2} 2\sqrt{\frac{16\pi}{5}} \Big( x \nu \mathcal{H}_{lm}^{\rm pp} +  x^6 \mathcal{H}_{lm}^{\rm AT}\Big) e^{-im\phi} \, .
\end{align}
where $R$ is the distance between the source and the observer.
Here, we present the dominant quadrupolar $(l=2,m=2)$ mode up to 2PN order since it is the most important one for data analysis of the gravitational waveform for circular orbits.
The point-particle contribution to this mode is given by \cite{Blanchet:2008je}
\begin{align}
    \mathcal{H}_{22}^{\rm pp} &= 1 + x \bigg\{ \frac{55 \nu }{42}-\frac{107  }{42} \bigg\} + x^{3/2} \bigg\{ 2\pi \bigg\} + x^2 \bigg\{ \frac{2047 \nu ^2}{1512}-\frac{1069 \nu}{216}-\frac{2173 }{1512} \bigg\}  +  \mathcal{O}\left(x^{5/2}\right)   \, ,
\end{align}
and the adiabatic tidal contribution is given by,
\begin{align}
    \mathcal{H}_{22}^{\rm AT} =&  \bigg\{\left( 12 \nu +3 \right)  \widetilde{\lambda}_{(+)} + 3 \delta \widetilde{\lambda}_{(-)}  \bigg\}  + x \bigg\{\left(\frac{45 \nu ^2}{7}-20 \nu +\frac{9}{2} \right) \widetilde{\lambda}_{(+)} +  \left( \frac{125 \nu }{7}+\frac{9}{2} \right) \delta \widetilde{\lambda}_{(-)}  \bigg\} \nonumber \\
    & + x^{3/2} \bigg\{ \left( 24 \nu +6 \right) \pi  \widetilde{\lambda}_{(+)} + 6 \pi \delta \widetilde{\lambda}_{(-)}  \bigg\} \\
    & + x^2 \bigg\{ \left(-\frac{274 \nu ^3}{21}-\frac{19367 \nu ^2}{168}-\frac{7211 \nu }{168}+\frac{1403}{56} \right) \widetilde{\lambda}_{(+)} +   \left( \frac{103 \nu ^2}{24}+\frac{1559 \nu }{56}+\frac{1403}{56} \right) \delta \widetilde{\lambda}_{(-)} \bigg\} +  \mathcal{O}\left(x^{5/2}\right)\nonumber
\end{align} 
We find agreement for $\mathcal{H}_{22}^{\rm AT}$ with the new results from Ref.~\cite{Dones:2024}.

\subsection{Phase of the waveform for circular orbits}
\label{sec_phase}
Now given that we have the expression of the flux from the equation \eqref{eq_flux} and the total energy of the system from Refs.~\cite{Henry:2019xhg,Mandal:2023lgy}, we can use the energy balance equation, 
\begin{align}
    \frac{dx}{dt} = - \frac{\mathcal{F}_E}{ dE/dx}   \, ;\quad\quad\quad\quad \frac{d \phi}{dt} = -\frac{5}{\nu} x^{3/2}  \, ,
\end{align}
to compute the phase evolution of the binary system $\phi(t)$. This can be then used  to compute half-phase of the emitted GW.
Using the flux balance equation, we first solve for $x(t)$, and then use the definition of the phase to obtain an expression for $\phi(t)$. This can then be expressed as,
\begin{align}
    \phi=\phi_0+\phi^{\rm pp}+\phi^{\rm AT}   \, ,
\end{align}
where, the $\phi_0$ is a integration constant which has to be fixed to a reference value, for example the phase at which the binary enters the detector band.
The point-particle sector up to 2PN order is given by \cite{Blanchet:1995fg,Will:1996zj},
\begin{align}
    \phi^{\rm pp} =& -\frac{1}{32 \nu x^{5/2}} \bigg[ 1+ x\bigg\{ \frac{3715}{1008}+ \frac{55\nu}{12} \bigg\} + x^{3/2}\bigg\{ -10\pi \bigg\} \nonumber\\
    &\quad\quad\quad\quad\quad\quad+ x^2 \bigg\{ \frac{15293365}{1016064}+ \frac{27145\nu}{1008}+ \frac{3085\nu^2}{144} \bigg\} +  \mathcal{O}\left(x^{5/2}\right) \bigg]   \, ,
\end{align}
and adiabatic tidal sector up to 2PN is given as,
\begin{align}
    \phi^{\rm AT} =& -\frac{3x^{5/2}}{16 \nu^2} \bigg[ \bigg\{ (1+22\nu) \widetilde{\lambda}_{(+)} + \delta \widetilde{\lambda}_{(-)} \bigg\} \nonumber\\
    &\quad\quad\quad\quad + x\bigg\{ \left( \frac{195}{56} + \frac{1595\nu}{14} + \frac{325\nu^2}{42} \right) \widetilde{\lambda}_{(+)} + \left( \frac{195}{56} + \frac{4415\nu}{168} \right) \delta \widetilde{\lambda}_{(-)} \bigg\} \nonumber\\
    &\quad\quad\quad\quad + x^{3/2} \bigg\{ -\frac{5\pi}{2}(1+22\nu) \widetilde{\lambda}_{(+)} -\frac{5\pi}{2} \delta \widetilde{\lambda}_{(-)} \bigg\} \nonumber\\
    &\quad\quad\quad\quad+x^2\bigg\{ \left( \frac{136190135}{9144576} + \frac{978554825\nu}{1524096} - \frac{281935\nu^2}{2016} +  5\nu^3 \right) \widetilde{\lambda}_{(+)} \nonumber\\ 
    & \quad\quad\quad\quad\quad\quad\quad + \left( \frac{136190135}{9144576} + \frac{213905\nu}{864} + \frac{1585\nu^2 }{432} \right) \delta \widetilde{\lambda}_{(-)} \bigg\} +  \mathcal{O}\left(x^{5/2}\right)
    \bigg]  \, .
\end{align}
The result presented in Ref. \cite{Henry:2020skiErr} agrees with the above expression of $\phi^{\rm AT}$.

We also compute the phase $\Psi$ of the emitted waveform in Fourier domain using the stationary phase approximation (SPA) \cite{Tichy:1999pv}. For the dominant $(l=2,m=2)$ quadrupolar mode, the frequency $f$ of the GW is twice the orbital frequency, and for convenience we define 
$v=(\pi G M f /c^3)^{1/3}$.
Then the phase $\Psi$ can be written as
\begin{align}
    \Psi_{\rm SPA} = 2 \pi f t_0 + \Psi_0+  \Psi_{\rm SPA}^{\rm pp}  + \Psi_{\rm SPA}^{\rm AT}   \, ,
\end{align}
where, $t_0$ and $\Psi_0$ (where we have absorbed the factor of $-\pi/4$) are integration constants. The point particle component up to 2PN is given by \cite{Blanchet:2023bwj},
\begin{align}
    \Psi_{\rm SPA}^{\rm pp} =& -\frac{3}{128 \nu v^5} \bigg[ 1 + v^2 \bigg\{ \frac{55 \nu }{9}+\frac{3715}{756} \bigg\} + v^{3} \bigg\{ 16 \pi  \bigg\} \nonumber\\
    &\quad\quad\quad\quad\quad + v^4\bigg\{ \frac{3085 \nu ^2}{72}+\frac{27145 \nu }{504}+\frac{15293365}{508032} \bigg\} +   \mathcal{O}\left(v^{5}\right)  \bigg]   \, ,
\end{align}
and the tidal contribution up to 2PN is given by,
\begin{align}
    \Psi_{\rm SPA}^{\rm AT} &= -\frac{9v^{5}}{16 \nu^2} \bigg[ \bigg\{ (1+22\nu) \widetilde{\lambda}_{(+)} + \delta \widetilde{\lambda}_{(-)} \bigg\} \nonumber\\
    &\quad\quad\quad\quad + v^2\bigg\{ \left( \frac{195}{112} + \frac{1595\nu}{28} + \frac{325\nu^2}{84} \right) \widetilde{\lambda}_{(+)} + \left( \frac{195}{112} + \frac{4415\nu}{336} \right) \delta \widetilde{\lambda}_{(-)} \bigg\} \nonumber\\
    &\quad\quad\quad\quad + v^{3} \bigg\{ -\pi (1+22\nu) \widetilde{\lambda}_{(+)} - \pi \delta \widetilde{\lambda}_{(-)} \bigg\} \nonumber\\
    &\quad\quad\quad\quad+v^4\bigg\{ \left( \frac{136190135}{27433728}+\frac{978554825 \nu }{4572288} -\frac{281935 \nu ^2}{6048} + \frac{5 \nu ^3}{3} \right) \widetilde{\lambda}_{(+)} \nonumber\\ 
    & \quad\quad\quad\quad\quad\quad\quad + \left( \frac{136190135}{27433728} +\frac{213905 \nu }{2592} +  \frac{1585 \nu ^2}{1296} \right) \delta \widetilde{\lambda}_{(-)} \bigg\} +  \mathcal{O}\left(v^{5}\right) \bigg]   \, .
\end{align}
The result presented in Ref. \cite{Henry:2020skiErr} agrees with the above expression of $\Psi_{\rm SPA}^{\rm AT}$.

\section{Conclusion}\label{sec_Conclusion} 

In this work, we have computed the source multipole moments, as well as the GW energy flux and angular momentum flux, for inspiraling compact binaries, taking into account tidal effects at next-to-next-to-leading PN order. Using the EFT framework and multi-loop Feynman diagram techniques, we calculated the potential and the stress-energy tensor up to 2PN. Specifically, we computed Feynman integrals up to two-loop order for the adiabatic sector and one-loop order for the point-particle sector, using dimensional regularization to determine the stress-energy tensor. Then we performed the explicit matching between the $d-$dimensional multipole-expanded effective theory and the stress-energy tensor to obtain the necessary multipole moments. 

A consistent gauge choice is typically required while matching the multipole moments, which involves applying the same coordinate shifts used in the conservative sector to the multipoles. In our approach, we propose to simultaneously vary the Lagrangian and the effective stress-energy tensor to eliminate accelerations and higher-order time derivatives. In doing so, we consistently modify the stress-energy tensor with the same shifts applied to the Lagrangian, thereby obtaining the source multipole moments for generic orbits in a gauge compatible with the conservative sector.
After including all contributions from the source multipole moments, as well as the hereditary effects, \textit{our computation based on the EFT approach provides the  gravitational wave flux 
for circular orbits at the 2PN order, as well as the quadrupolar GW mode amplitude and GW phases at the same order.}
In addition, we performed rigorous self-consistency checks on our computation of the stress-energy tensor up to the 2PN order by constructing multiple moment equations derived from the conservation condition $\partial_\mu \mathcal{T}^{\mu\nu}=0$. These checks required the calculation of the complete stress-energy tensor up to 2.5PN order.

The results presented in this article, particularly that of the gravitational energy
flux, were used to identify a mistake in the previous result~\cite{Henry:2020ski}, which after correction~\cite{Henry:2020skiErr} agree with ours.
Several waveform models~\cite{Williams:2024twp,Colleoni:2023czp,Gamba:2023mww,Narikawa:2023deu,Tissino:2022thn,Abac:2023ujg} and analyses using them~\cite{LIGOScientific:2024elc,Koehn:2024set,Golomb:2024mmt,Narikawa:2022saj,Kuan:2022etu}
relied on the results in Ref.~\cite{Henry:2020ski}.
We expect that the results presented in this article will be used where appropriate, thereby improving the reliability of the models.

The outcomes of the present study pave the way for improved accuracy in modeling gravitational waves generated by tidally deformed binary systems.  In this work, we have focused on the adiabatic quadrupolar tide on the compact object. However, our framework and automatic computational techniques can be extended to obtain other higher-order corrections, which are crucial for improving the waveform templates \cite{
Samajdar:2018dcx,Samajdar:2019ulq} and are essential for gravitational wave detection and interpretation, especially with the upcoming observatories. One can easily include higher-order multipolar adiabatic tides as studied in Refs.~\cite{Bini:2012gu,Thorne:1984mz,Zhang:1986cpa,Damour:1990pi,Binnington:2009bb,Damour:2009wj,Damour:2009vw,Henry:2019xhg} in our framework. Other types of interesting physical effects can also be included, such as dynamical tides due to the different oscillation modes of a NS. The quadrupolar oscillation modes are studied in the conservative sector in Refs.~\cite{Steinhoff:2016rfi,Gupta:2020lnv,Gupta:2023oyy,Mandal:2023hqa,Mandal:2023lgy}, but have not yet been included to the same PN order in the radiative sector. We will leave this for future analysis. 


\subsection*{Acknowledgements}
RP thanks Quentin Henry for cross-checks and discussions during the course of this project. 
R.P.’s research is funded by the Deutsche Forschungsgemeinschaft (DFG, German Research Foundation), Projektnummer 417533893/GRK2575 “Rethinking Quantum Field Theory”.

\appendix
\addtocontents{toc}{\protect\setcounter{tocdepth}{1}}

\section{Point particle effective Lagrangian and Multipoles}
\label{app_non_tidal}
In this section we provide the necessary equations for the point-particle sector.
\subsection{Effective Lagrangian}
\label{app_non_tidal_lag}
The Lagrangian for the point-particle sector is given by,
\begin{align}
    \widetilde{\mathcal{L}}^{\rm pp} = \widetilde{\mathcal{L}}^{\rm pp}_{\rm 0PN} + \widetilde{\mathcal{L}}^{\rm pp}_{\rm 1PN} + \widetilde{\mathcal{L}}^{\rm pp}_{\rm 2PN} + \mathcal{O}\left(\frac{1}{c^5}\right)   \, ,
\end{align}
where, the individual components at different PN orders are given as
\begin{align}
    \widetilde{\mathcal{L}}_{\rm 0PN} =& \frac{\widetilde{v}^2}{2}+\frac{1}{\widetilde{r}}   \, ;\\
    \widetilde{\mathcal{L}}_{\rm 1PN} =&  \left(\frac{1}{8}-\frac{3 \nu }{8}\right) \widetilde{v}^4 + \frac{1}{\widetilde{r}} \left(\frac{\nu  (\textbf{n}\cdot\widetilde{\textbf{v}})^2}{2}+\left(\frac{\nu }{2}+\frac{3}{2}\right) \widetilde{v}^2\right)-\frac{1}{2\widetilde{r}^2}   \, ;\\
    \widetilde{\mathcal{L}}_{\rm 2PN} =&  \left(\frac{13 \nu ^2}{16}-\frac{7 \nu }{16}+\frac{1}{16}\right) \widetilde{v}^6 +\frac{1}{\widetilde{r}} \left(\frac{3 \nu ^2 (\textbf{n}\cdot\widetilde{\textbf{v}})^4}{8}+\left(\nu -\frac{5 \nu ^2}{4}\right) (\textbf{n}\cdot\widetilde{\textbf{v}})^2 \widetilde{v}^2+\left(-\frac{9 \nu ^2}{8}-2 \nu +\frac{7}{8}\right) \widetilde{v}^4\right) \nonumber\\
    &+
    \frac{1}{\widetilde{r}^2} \left(\left(\frac{3 \nu ^2}{2}+\frac{3 \nu }{2}+\frac{1}{2}\right) (\textbf{n}\cdot\widetilde{\textbf{v}})^2+\left(\frac{\nu ^2}{2}+\frac{7}{4}\right) \widetilde{v}^2\right)+\left(\frac{\nu }{4}+\frac{1}{2}\right) \frac{1}{\widetilde{r}^3}   \, ;
\end{align}
This Lagrangian is given in the same gauge as in Ref.~\cite{Mandal:2022nty} and is related to it by a Legendre transformation. 
We also derive the relation for $r$ in terms of the PN parameter $x$ for circular orbits. It is given by 
\begin{align}
\label{eq_cir_orb_pp}
    \frac{1}{\widetilde{r}} = x + x^2 \left( 1-\frac{\nu}{3} \right) + x^3 \left( 1-\frac{43\nu}{24} \right) +  \mathcal{O}\left(x^{7/2}\right)  \, .
\end{align}

\subsection{Multipole moments}
\label{app_pp_multimom}
Different multipole moments that are derived using a 2PN stress-energy tensor, for the point-particle sector, are given by,
\begin{align}
    \Big(\widetilde{\mathcal{I}}^{\rm pp}\Big)^{ij} =& \Big(\widetilde{\mathcal{I}}_{\rm 0PN}^{\rm pp}\Big)^{ij}+\Big(\widetilde{\mathcal{I}}_{\rm 1PN}^{\rm pp}\Big)^{ij}+\Big(\widetilde{\mathcal{I}}_{\rm 2PN}^{\rm pp}\Big)^{ij}   \, ;\\
    \Big(\widetilde{\mathcal{I}}^{\rm pp}\Big)^{ijk} =& \Big(\widetilde{\mathcal{I}}_{\rm 0PN}^{\rm pp}\Big)^{ijk}+\Big(\widetilde{\mathcal{I}}_{\rm 1PN}^{\rm pp}\Big)^{ijk}   \, ;\\
    \Big(\widetilde{\mathcal{I}}^{\rm pp}\Big)^{ijkl} =& \Big(\widetilde{\mathcal{I}}_{\rm 0PN}^{\rm pp}\Big)^{ijkl}  \, ,
\end{align}
where the individual components for each PN order are given as,
\begin{align}
    \Big(\widetilde{\mathcal{I}}_{\rm 0PN}^{\rm pp}\Big)^{ij} = & \Big[ \textbf{n}^i\textbf{n}^j \Big]_{ \texttt{\tiny STF}} \widetilde{r}^2    \, ;\\
    \Big(\widetilde{\mathcal{I}}_{\rm 1PN}^{\rm pp}\Big)^{ij} = &  
    \Big[ \textbf{n}^i\textbf{n}^j \Big]_{ \texttt{\tiny STF}} \bigg\{ \left(\frac{29}{42}-\frac{29 \nu }{14}\right) \widetilde{r}^2 \widetilde{v}^2+\left(\frac{8 \nu }{7}-\frac{5}{7}\right) r \bigg\}  
    + \Big[ \textbf{n}^i\widetilde{\textbf{v}}^j \Big]_{ \texttt{\tiny STF}}\bigg\{ (\textbf{n}\cdot\widetilde{\textbf{v}}) \widetilde{r}^2 \left(\frac{12 \nu }{7}-\frac{4}{7}\right) \bigg\} \nonumber \\
    & + \Big[ \widetilde{\textbf{v}}^i\widetilde{\textbf{v}}^j \Big]_{ \texttt{\tiny STF}} \bigg\{ \left(\frac{11}{21}-\frac{11 \nu }{7}\right) \widetilde{r}^2 \bigg\}   \, ;\\
    \Big(\widetilde{\mathcal{I}}_{\rm 2PN}^{\rm pp}\Big)^{ij} = & 
    \Big[ \textbf{n}^i\textbf{n}^j \Big]_{ \texttt{\tiny STF}} \bigg\{ \left(\frac{337 \nu ^2}{252}-\frac{71 \nu }{126}-\frac{355}{252}\right) + \left(-\frac{1273 \nu ^2}{756}+\frac{359 \nu }{378}-\frac{131}{756}\right) (\textbf{n}\cdot\widetilde{\textbf{v}})^2 \widetilde{r} \nonumber\\
    & \quad\quad\quad\quad +\left(\frac{3545 \nu ^2}{504}-\frac{1835 \nu }{504}+\frac{253}{504}\right) \widetilde{r}^2 \widetilde{v}^4+\left(-\frac{4883 \nu ^2}{756}-\frac{2879 \nu }{378}+\frac{2021}{756}\right) \widetilde{r} \widetilde{v}^2 \bigg\}  \nonumber\\
    & + \Big[ \textbf{n}^i\widetilde{\textbf{v}}^j \Big]_{ \texttt{\tiny STF}} \bigg\{  \left(-\frac{418 \nu ^2}{63}+\frac{202 \nu }{63}-\frac{26}{63}\right) (\textbf{n}\cdot\widetilde{\textbf{v}}) \widetilde{r}^2 \widetilde{v}^2+\left(\frac{209 \nu ^2}{54}+\frac{800 \nu }{189}-\frac{155}{54}\right) (\textbf{n}\cdot\widetilde{\textbf{v}}) \widetilde{r} \bigg\} \nonumber \nonumber\\
    & + \Big[ \widetilde{\textbf{v}}^i\widetilde{\textbf{v}}^j \Big]_{ \texttt{\tiny STF}} \bigg\{ \left(\frac{25 \nu ^2}{63}-\frac{25 \nu }{63}+\frac{5}{63}\right) (\textbf{n}\cdot\widetilde{\textbf{v}})^2 \widetilde{r}^2+\left(\frac{733 \nu ^2}{126}-\frac{337 \nu }{126}+\frac{41}{126}\right) \widetilde{r}^2 \widetilde{v}^2 \nonumber\\
    & \quad\quad\quad\quad +\left(-\frac{985 \nu ^2}{189}-\frac{335 \nu }{189}+\frac{106}{27}\right) \widetilde{r} \bigg\}   \, ;
\end{align}

\begin{align}
    \Big(\widetilde{\mathcal{I}}_{\rm 0PN}^{\rm pp}\Big)^{ijk} =& \Big[ \textbf{n}^i\textbf{n}^j\textbf{n}^k \Big]_{ \texttt{\tiny STF}}  \widetilde{r}^3   \, ;\\
    \Big(\widetilde{\mathcal{I}}_{\rm 1PN}^{\rm pp}\Big)^{ijk} =& \Big[ \textbf{n}^i\textbf{n}^j\textbf{n}^k \Big]_{ \texttt{\tiny STF}}  \bigg\{ \left(\frac{5}{6}-\frac{19 \nu }{6}\right) \widetilde{r}^3 \widetilde{v}^2+\left(\frac{13 \nu }{6}-\frac{5}{6}\right) \widetilde{r}^2 \bigg\}   
    + \Big[ \textbf{n}^i\textbf{n}^j\widetilde{\textbf{v}}^k \Big]_{ \texttt{\tiny STF}} \bigg\{ (2 \nu -1) (\textbf{n}\cdot\widetilde{\textbf{v}}) \widetilde{r}^3 \bigg\} \nonumber\\
    &+ \Big[ \textbf{n}^i\widetilde{\textbf{v}}^j\widetilde{\textbf{v}}^k \Big]_{ \texttt{\tiny STF}} \bigg\{ (1-2 \nu ) \widetilde{r}^3 \bigg\}   \, ;
\end{align}

\begin{align}
    \Big(\widetilde{\mathcal{I}}_{\rm 0PN}^{\rm pp}\Big)^{ijkl} =& \Big[ \textbf{n}^i\textbf{n}^j\textbf{n}^k\textbf{n}^l \Big]_{ \texttt{\tiny STF}} \bigg\{ (1-3 \nu ) \widetilde{r}^4 \bigg\}  \, .
\end{align}

For the current multipole moments, we find 
\begin{align}
    \Big(\widetilde{\mathcal{J}}^{\rm pp}\Big)^{ij} =& \Big(\widetilde{\mathcal{J}}_{\rm 0PN}^{\rm pp}\Big)^{ij}+\Big(\widetilde{\mathcal{J}}_{\rm 1PN}^{\rm pp}\Big)^{ij}  \, ;\\
    \Big(\widetilde{\mathcal{J}}^{\rm pp}\Big)^{ijk} =& \Big(\widetilde{\mathcal{J}}_{\rm 0PN}^{\rm pp}\Big)^{ijk}  \, ,
\end{align}
where,
\begin{align}
    \Big(\widetilde{\mathcal{J}}_{\rm 0PN}^{\rm pp}\Big)^{ij} =& \Big[ \epsilon^{imn}\textbf{n}^j\textbf{n}^m\widetilde{\textbf{v}}^n \Big]_{ \texttt{\tiny STF}} \bigg\{  -\widetilde{r}^2  \bigg\}   \, ;\\
    \Big(\widetilde{\mathcal{J}}_{\rm 1PN}^{\rm pp}\Big)^{ij} =& \Big[ \epsilon^{imn}\textbf{n}^j\textbf{n}^m\widetilde{\textbf{v}}^n \Big]_{ \texttt{\tiny STF}} \bigg\{\left(\frac{17 \nu }{7}-\frac{13}{28}\right) \widetilde{r}^2 \widetilde{v}^2+\left(-\frac{15 \nu }{7}-\frac{27}{14}\right) \widetilde{r} \bigg\} \nonumber\\
    &+ \Big[ \epsilon^{imn}\widetilde{\textbf{v}}^j\textbf{n}^m\widetilde{\textbf{v}}^n \Big]_{ \texttt{\tiny STF}} \bigg\{ \left(\frac{5 \nu }{14}-\frac{5}{28}\right) (\textbf{n}\cdot\widetilde{\textbf{v}}) \widetilde{r}^2 \bigg\}   \, ;
\end{align}

\begin{align}
    \Big(\widetilde{\mathcal{J}}_{\rm 0PN}^{\rm pp}\Big)^{ijk} =& \Big[ \epsilon^{imn}\textbf{n}^j\textbf{n}^k\textbf{n}^m\widetilde{\textbf{v}}^n \Big]_{ \texttt{\tiny STF}} \bigg\{ (1-3 \nu ) \widetilde{r}^3 \bigg\}  \, .
\end{align}

\section{Tidal Effective Lagrangians}
\label{app_tidal}

The Lagrangian of the tidal sector is obtained by removing higher-order time derivatives using coordinate shifts as prescribed in section \ref{sec_com_algo}. This Lagrangian is in a different gauge as the Hamiltonian presented in our previous work \cite{Mandal:2023lgy}, since that was computed by taking the adiabatic limit of the dynamic Hamiltonian. 
We decompose the Lagrangian as,
\begin{align}
    \widetilde{\mathcal{L}}^{\rm AT} = \widetilde{\mathcal{L}}^{\rm AT}_{\rm 0PN} + \widetilde{\mathcal{L}}^{\rm AT}_{\rm 1PN} + \widetilde{\mathcal{L}}^{\rm AT}_{\rm 2PN} + \mathcal{O}\left(\frac{1}{c^5}\right)  \, ,
\end{align}
where, the indivudual contribution to different PN orders is given by
\begin{align}
    \widetilde{\mathcal{L}}_{\rm 0PN}^{\rm AT} =& \lambda_{(1)} \bigg\{ \frac{3}{2 q \widetilde{r}^6} \bigg\} +(1\leftrightarrow 2)  \, ;\\
    \widetilde{\mathcal{L}}_{\rm 1PN}^{\rm AT} =&  \lambda_{(1)} \bigg\{ \frac{1}{\widetilde{r}^6} \left(\frac{\nu  \widetilde{v}^2}{4}-\frac{3 \nu  (\textbf{n}\cdot\widetilde{\textbf{v}})^2}{2}\right)+\frac{7 \nu }{2\widetilde{r}^7} \nonumber \\
    &\quad\quad + \frac{1}{q} \left[\frac{1}{\widetilde{r}^6} \left(\left(3 \nu -\frac{9}{2}\right) (\textbf{n}\cdot\widetilde{\textbf{v}})^2+\left(\nu +\frac{15}{4}\right) \widetilde{v}^2\right)+\left(\frac{7 \nu }{2}-\frac{21}{2}\right) \frac{1}{\widetilde{r}^7}\right]\bigg\} +(1\leftrightarrow 2)  \, ;\\
    \widetilde{\mathcal{L}}_{\rm 2PN} =&  \lambda_{(1)} \bigg\{
    \frac{1}{\widetilde{r}^6} \left(\left(15 \nu -6 \nu ^2\right) (\textbf{n}\cdot\widetilde{\textbf{v}})^4+\left(\frac{9 \nu ^2}{2}-\frac{33 \nu }{4}\right) (\textbf{n}\cdot\widetilde{\textbf{v}})^2 \widetilde{v}^2+\left(\frac{7 \nu }{16}-\frac{3 \nu ^2}{8}\right) \widetilde{v}^4\right)\nonumber\\
    &\quad\quad+\frac{1}{\widetilde{r}^7} \left(\left(8 \nu ^2-\frac{201 \nu }{2}\right) (\textbf{n}\cdot\widetilde{\textbf{v}})^2+\left(2 \nu ^2+\frac{129 \nu }{4}\right) \widetilde{v}^2\right)-\frac{1}{28} 961 \nu  \frac{1}{\widetilde{r}^8}\nonumber\\
    &\quad\quad+ \frac{1}{q} \bigg[
    \frac{1}{\widetilde{r}^6} \bigg(\left(3 \nu ^2-12 \nu +\frac{9}{2}\right) (\textbf{n}\cdot\widetilde{\textbf{v}})^4+\left(-\frac{27 \nu ^2}{4}+24 \nu -\frac{45}{4}\right) (\textbf{n}\cdot\widetilde{\textbf{v}})^2 \widetilde{v}^2  \nonumber\\
    &\quad\quad+\left(-\frac{33 \nu ^2}{16}-\frac{11 \nu }{2}+\frac{105}{16}\right) \widetilde{v}^4\bigg)+\frac{1}{\widetilde{r}^7} \bigg(\left(\frac{55 \nu ^2}{2}-\frac{1183 \nu }{8}+\frac{27}{2}\right) (\textbf{n}\cdot\widetilde{\textbf{v}})^2 \nonumber\\
    & \quad\quad +\left(\frac{7 \nu ^2}{2}+\frac{247 \nu }{8}-\frac{45}{4}\right) \widetilde{v}^2\bigg)+ \left(\frac{165}{4}-\frac{239 \nu }{8}\right) \frac{1}{\widetilde{r}^8}
    \bigg]
    \bigg\} +(1\leftrightarrow 2)  \, .
\end{align}
Here we also derive the relation between $r$ and the frequency of the circular orbit. It is given by 
\begin{align}
\label{eq_cir_orb_AT}
    \frac{1}{\widetilde{r}} =& x^6 \bigg\{ 6 \lambda_{(+)} \bigg\} + x^7 \bigg\{  \left( -\frac{29}{2}+3 \nu \right)\lambda_{(+)} -  \frac{17}{2} \delta\lambda_{(-)}\bigg\}  \nonumber\\
    &+ x^8 \bigg\{  \left( -\frac{2563}{56}+ \frac{263 \nu }{14} +15 \nu ^2 \right) \lambda_{(+)} +   \left( -\frac{1891}{56}+ \frac{13 \nu }{2} \right) \delta\lambda_{(-)} \bigg\}  +  \mathcal{O}\left(x^{17/2}\right)   \, .
\end{align}

\bibliographystyle{JHEP}
\bibliography{biblio}

\end{document}